\documentclass[aps,prd,groupedaddress,preprint,eqsecnum,nofootinbib,showpacs]{revtex4}
\usepackage{graphicx,epsf,amssymb,amsbsy,amsfonts,amssymb,amsmath}
\usepackage[dvips]{color}

\newcommand{\beq}{\begin{equation}}
\newcommand{\eeq}{\end{equation}}
\newcommand{\be}{\begin{equation}}
\newcommand{\ee}{\end{equation}}
\newcommand{\beqa}{\begin{eqnarray}}
\newcommand{\eeqa}{\end{eqnarray}}
\newcommand{\beqar}{\begin{eqnarray*}}
\newcommand{\eeqar}{\end{eqnarray*}}
\newcommand{\bea}{\begin{eqnarray}}
\newcommand{\eea}{\end{eqnarray}}

\newcommand{\dash}{\text{-}}

\newcommand{\cn}{{\cal N}}

\newcommand{\nn}{\nonumber}

\def\be{\begin{equation}}
\def\ee{\end{equation}}
\def\bea{\begin{eqnarray}}
\def\eea{\end{eqnarray}}
\def\ba{\begin{array}}
\def\ea{\end{array}}
\def\bd{\begin{displaymath}}
\def\ed{\end{displaymath}}
\def\eqn#1{eq.~(\ref{#1})}

\def\fig#1{fig.~\ref{#1}}
\def\sect#1{sect.~\ref{#1}}

\def\n{{\tilde n}}

\def\>{\rangle}
\def\<{\langle}
\def\Dsl{D \hskip-.6em \raise1pt\hbox{$ / $ } }
\def\to{\rightarrow}

\def\NeqFour{{{\cal N}=4}}

\def\NeqEight{{{\cal N}=8}}
\def\tree{{\rm tree}}
\def\scalar{{\rm scalar}}
\def\f{{\tilde f}}

\newcommand{\ie}{{\it i.e.}\ }
\newcommand{\Tr}{\textrm{Tr}}

\newcommand{\tn}{\tilde{n}}

\numberwithin{equation}{section}

\begin{document}

\setlength{\unitlength}{1mm}

\vspace{4mm}

\begin{flushright}
PUPT-2335 \\
UCLA/TEP/10/103 \\
\end{flushright}


\title{Gravity as the Square of Gauge Theory}

\author{Zvi Bern${}^{a}$,
     Tristan Dennen${}^{a}$,
     Yu-tin Huang${}^{a}$,
     Michael Kiermaier$^{b}$ \\[10mm]
{{${}^{a}${\it Department of Physics and Astronomy}\\
         {\it UCLA, Los Angeles, CA}\\
         {\it 90095-1547, USA}}\\[5mm]
{${}^{b}${\it Joseph Henry Laboratories}\\
{\it Princeton University}\\
{\it Princeton, NJ 08544, USA}}\\[5mm]
$\null$\\
}
}

\vskip .3truecm

\begin{abstract}
\vskip .25truecm
\noindent
We explore consequences of the recently discovered duality between
color and kinematics, which states that kinematic numerators in a
diagrammatic expansion of gauge-theory amplitudes can be arranged to
satisfy Jacobi-like identities in one-to-one correspondence to the
associated color factors.  Using on-shell recursion relations, we give
a field-theory proof showing that the duality implies that
diagrammatic numerators in gravity are just the product of two
corresponding gauge-theory numerators, as previously conjectured.
These squaring relations express gravity amplitudes in terms of
gauge-theory ingredients, and are a recasting of the Kawai, Lewellen
and Tye relations. Assuming that numerators of loop amplitudes can be
arranged to satisfy the duality, our tree-level proof
immediately carries over to loop level via
the unitarity method.  We then present a Yang-Mills Lagrangian whose
diagrams through five points manifestly satisfy the duality between
color and kinematics. The existence of such Lagrangians suggests
 that the duality also extends to loop amplitudes, as confirmed
 at two and three loops in a concurrent paper.  By ``squaring'' the novel Yang-Mills
Lagrangian we immediately obtain its gravity counterpart.  We outline
the general structure of these Lagrangians for higher points.  We also
write down various new representations of gauge-theory and gravity
amplitudes that follow from the duality between color and
kinematics.
\end{abstract}

\pacs{04.65.+e, 11.15.Bt, 11.30.Pb, 11.55.Bq \hspace{1cm}}

\maketitle

\tableofcontents

\section{Introduction}

A key lesson from studies of scattering amplitudes is that weakly
coupled gauge and gravity theories have a far simpler and richer
structure than is evident from their usual Lagrangians.  A striking
example of this is Witten's remarkable conjecture that scattering
amplitudes in twistor space are supported on curves of a degree
controlled by their helicity and loop
order~\cite{WittenTopologicalString}.  At weak coupling another
remarkable structure visible in on-shell tree amplitudes are the
Kawai-Lewellen-Tye (KLT) relations, which express gravity tree-level
amplitudes as sums of products of gauge-theory
amplitudes~\cite{KLT,GravityReview}.  These relations were originally
formulated in string theory, but hold just as well in field theory.
In fact, in many cases, they hold even when no string theory lives
above the field theory~\cite{HeteroticKLT}.

The KLT relations have recently been recast into a much simpler form
in terms of numerators of diagrams with only
three-point vertices.
In the
new representation the diagrammatic numerators in gravity are simply a
product of two corresponding gauge-theory
numerators~\cite{BCJ}.  Underlying these numerator ``squaring
relations'' is a newly discovered duality between kinematic
numerators of gauge theory and their associated color factors, by
Carrasco, Johansson and one of the authors (BCJ).  The BCJ duality
states that gauge-theory amplitudes can be arranged into a form where
diagrammatic numerators satisfy a set of identities in one-to-one
correspondence to the Jacobi identities obeyed by color factors.  The
duality appears to hold in large classes of theories including pure
Yang-Mills theory and $\NeqFour$ super-Yang-Mills.  BCJ conjectured
that the numerators of gravity diagrams are simply the product of two
corresponding gauge-theory numerators that satisfy the duality.
These squaring relations were verified in ref.~\cite{BCJ} at tree level up
to eight points.  Interestingly, the duality also leads to a set of
non-trivial relations between gauge-theory amplitudes~\cite{BCJ},
which are now well understood in string theory~\cite{Bjerrum1}.  The
numerator duality relations have also been understood from the vantage
point of string theory~\cite{Mafra,Tye,Bjerrum2}. In particular, the
heterotic string offers important insight into these relations,
because of the parallel treatment of color and kinematics~\cite{Tye}.

In this paper we describe two complementary approaches to developing a
field-theory understanding of the duality between color and
kinematics, and its relation to gravity as two copies of gauge theory.
In the first approach we use BCFW on-shell recursion relations to
prove that the squaring relations are satisfied if the numerators of
gauge-theory diagrams satisfy the BCJ duality.  Our proof is
inductive, starting with three points where it is simple to verify the
double-copy property for candidate gravity theories.  For higher
points, we apply the BCFW recursion relations to gauge-theory
amplitudes whose numerators are arranged to satisfy the duality.  The
BCFW recursion relations, however, in general do not respect the
duality. This requires us to apply a ``generalized gauge
transformation'' to rearrange terms in the recursion relation in a way
that restores the duality, which is a key ingredient in our proof.
These generalized gauge transformations correspond to the most general
rearrangements of amplitude numerators that do not alter their
values. (Such transformations need not correspond to gauge
transformations in the traditional sense.)  To apply this to gravity
we make use of the fact that BCFW recursion relations for
color-dressed gauge-theory amplitudes~\cite{BCFW} are closely related
to the gravity ones~\cite{BCFWGravity}.  By also applying a
generalized gauge transformation to the BCFW recursion relation in
gravity, we show that the squaring relations indeed reproduce the
gravity amplitude correctly.

The generalized gauge invariance contains an enormous freedom in
rearranging amplitudes, and for some rearrangements the squaring
relations between gravity and gauge theory hold.  Such generalizations
of the squaring relations at five points were discussed in
refs.~\cite{Tye,Bjerrum2}.  Here we present an all-$n$ generalization
of the squaring relations given in an asymmetric form, in which only
one of the two sets of gauge-theory numerators is required to satisfy
the BCJ duality.\footnote{H.~Johansson independently realized that
  only one set of numerators needs to satisfy the BCJ duality to
  obtain gravity from the numerator squaring relation; see
  ref.~\cite{BCJLoops} for a non-trivial loop-level application.}

The unitarity method~\cite{UnitarityMethod} immediately implies that
gravity loop amplitudes must have the double-copy property, if the
corresponding gauge-theory loop amplitudes can be put in a form that
satisfies the BCJ duality, as does indeed appear to be the
case~\cite{BCJLoops}.  The squaring relations then apply to gravity
numerators for {\em any} value of loop momenta, {\it i.e.} with no cut
conditions applied.  This is to be contrasted with the KLT relations,
which are valid only at tree level, and can be applied at loop level
only on unitarity cuts that decompose loop amplitudes into tree
amplitudes~\cite{BDDPR}.  The KLT relations take a different functional form for
every cut of a given amplitude, depending on the precise
tree-amplitude factors involved in the cut.  The squaring relations,
on the other hand, take a simple universal form for any choice of loop
momenta.

In our second approach to understand the color-kinematics duality, we
use a more traditional Lagrangian viewpoint.  A natural question is:
what Lagrangian generates diagrams that automatically satisfy the BCJ
duality?  We shall describe such a Lagrangian here, and present its
explicit form up to five points, leaving the question of the more
complicated explicit higher-point forms to the future.  We have also
worked out the six-point Lagrangian and outline its structure, and
make comments about the all-orders form of the Lagrangian.  We find
that a covariant Lagrangian whose diagrams satisfy the duality is
necessarily non-local.  We can make this Lagrangian local by
introducing auxiliary fields.  Remarkably we find that, at least
through six points, the Lagrangian differs from ordinary Feynman gauge
simply by the addition of an appropriate zero, namely terms that
vanish by the color Jacobi identity.  Although the additional terms
vanish when summed, they appear in diagrams in just the right way so
that the BCJ duality is satisfied.  Based on the structures we find,
it seems likely that any covariant Lagrangian where diagrams with an arbitrary
number of external legs satisfy the duality must have an infinite
number of interactions.

In ref.~\cite{BernGrant}, the problem was posed of how to construct a
Lagrangian that reflects the double-copy property of gravity. That
reference carried out some initial steps, showing that one can
factorize the graviton indices into ``left'' and ``right'' classes
consistent with the factorization observed in the KLT relations.  (See
also ref.~\cite{Siegel}.)  Unfortunately, beyond three points the
relationship of the constructed gravity Lagrangian to gauge theory was
rather obscure.  As it turns out, a key ingredient was missing: the
duality between color and kinematics, which was discovered much
later~\cite{BCJ}.  Using the modified local version of the
gauge-theory Lagrangian whose Feynman diagrams respect the BCJ
duality, we construct a Lagrangian for gravity valid through five
points, as a double copy of the gauge-theory one.  The likely
appearance of an infinite number of interactions in the modified
gauge-theory Lagrangian is perhaps natural, because we expect any
covariant gravity Lagrangian to also have an infinite number of terms.

We present a simple application of the BCJ duality.  Since the BCJ
duality states that diagrammatic numerators have the same algebraic
structure as color factors, we can immediately make use of different
known color representations of amplitudes to write dual formulas where
color and kinematic numerators are swapped.  In particular, Del Duca,
Dixon and Maltoni~\cite{DDMColor} have given a color decomposition of
tree amplitudes using adjoint-representation color matrices.  They
derived this color decomposition using the color Jacobi identity and
Kleiss-Kuijf relations~\cite{Kleiss}.  By swapping color and numerator
factors in their derivation, we immediately obtain novel forms of both
gauge-theory and gravity tree amplitudes.

This paper is organized as follows.  In \sect{BCJDualityReviewSection}
we review BCJ duality.  We discuss invariances of the amplitudes in
sect.~\ref{InvarianceSection}, and also present a new, asymmetric form
of the squaring relations.  We give our proof that the kinematic
diagrammatic numerators for gravity are double copies of the
gauge-theory ones in \sect{SquaringSection}.  Then, in
\sect{LagrangianSection}, we turn to the question of constructing a
gauge-theory Lagrangian that generates Feynman diagrams that respect
the BCJ duality.  A few simple implications of BCJ duality are given
in \sect{SimpleImplicationsSection}.

\section{Review of BCJ duality}
\label{BCJDualityReviewSection}

\begin{figure}[t]
\centerline{\epsfxsize 5 truein \epsfbox{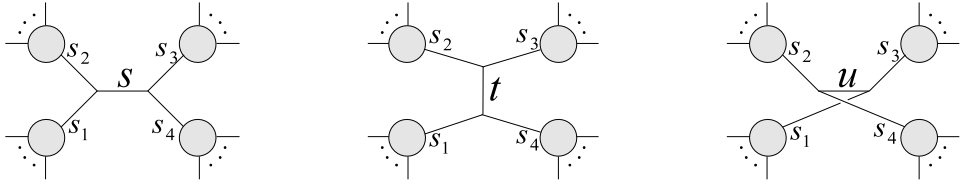}}
\caption[a]{\small A Jacobi relation between color factors of
  diagrams.  According to the BCJ duality the diagrammatic numerators
  of amplitudes can be arranged
  in a way that they satisfy relations in one-to-one correspondence to
  the color Jacobi identities.   }
\label{JacobiFigure}
\end{figure}

\subsection{General Considerations}

Consider a gauge-theory amplitude, which we write in a diagrammatic
form,
\begin{equation}
{1\over g^{n-2}}\,{\cal A}^\tree_n(1,2,3,\ldots,n)\,=\sum_{{\rm diags.}\, i}
                 {n_i c_i \over \prod_{\alpha_i} s_{\alpha_i}}\,,
\label{Anrep}
\end{equation}
where the sum runs over all diagrams $i$ with only three-point
vertices, the $c_i$ are color factors, the $n_i$ are kinematic
numerators, and the $s_{\alpha_i}$ are the inverse propagators
associated with the channels $\alpha_i$ of the diagram $i$.  Any
gauge-theory amplitude can be put into this form by replacing contact
terms with numerator factors canceling propagators, {\it i.e.,}
$s_\alpha/s_\alpha$ and assigning the contribution to the proper
diagram according to the color factor.  The value of the color
coefficient $c_i$ of each term is obtained from the diagram $i$ by
dressing each three-point vertex with a structure constant $\f^{abc}$, where
\begin{equation}
\f^{abc} \equiv i\sqrt{2} f^{abc} = \Tr\bigl( [T^a,T^b] \, T^c \bigr),
\label{ftildedef}
\end{equation}
and dressing each internal line with $\delta^{ab}$.

A key property of the $\f^{abc}$ is that they satisfy the Jacobi
identity.
Consider, for example, the color factors of the three diagrams
illustrated in \fig{JacobiFigure}.
They take the schematic form,
\begin{eqnarray}
 && c_s \equiv \ldots \f^{a_1 a_2 b} \f^{b a_3 a_4}\ldots \,, \hskip 1 cm
 c_t \equiv \ldots \f^{  a_1 a_4 b}\f^{b a_2 a_3 }  \ldots \,, \hskip 1 cm
  c_u \equiv \ldots \f^{ a_1 a_3  b} \f^{b a_4 a_2 }  \ldots \,, \nn \\
\end{eqnarray}
where the `$\dots$' signify factors common to all
three diagrams.  The
color factors then, of course, satisfy the Jacobi identity
\begin{equation}
 c_s+c_t+c_u=0\,.
\label{jacobic}
\end{equation}
Here we have chosen a sign convention\footnote{In any given Jacobi
  relation the relative signs are arbitrary since they always can be
  moved between color factors and kinematic numerators.}  that
differs from ref.~\cite{BCJ}.

The BCJ conjecture states that numerators $n_i$ can always be found
that satisfy Jacobi relations in one-to-one correspondence with the
color Jacobi identities,
\begin{equation}
c_i + c_j + c_k = 0 \,  \hskip 1cm \Rightarrow
\hskip 1 cm
n_i + n_j + n_k = 0\,,
\label{jacobin}
\end{equation}
where $i,j$ and $k$ label diagrams whose color factors are related by
a Jacobi identity.  (In general the relative signs between the color
factors in all Jacobi identities cannot be taken to be globally
positive, but according to the BCJ conjecture the relative signs
always match between the color and kinematic identities.)  In addition,
BCJ duality also requires that the $n_i$ satisfy the same
self-anti-symmetry relations as the $c_i$.  That is, if a color factor
is anti-symmetric under an interchange of two legs, the corresponding
numerator satisfies the same antisymmetry relations,
\begin{equation}
c_i \rightarrow -c_i \hskip 1 cm  \Rightarrow
\hskip 1 cm
 n_i \rightarrow -n_i\,.
\label{SelfAntisym}
\end{equation}
We note that when the color-ordered partial amplitudes are expressed
in terms of numerators satisfying these self-anti-symmetry relations,
they automatically satisfy the Kleiss-Kuijf
relations~\cite{Kleiss, DDMColor} between color-ordered partial
amplitudes~\cite{BCJ}.

\subsection{Five-point example and generalized Jacobi-like structures}

Consider the five-point case as a simple example, discussed already in
some detail from various viewpoints in refs.~\cite{BCJ,Mafra,Tye,Bjerrum2}.
At five points there are 15 numerators and 9 independent duality
relations, leaving 6 numerators.  Of these remaining numerators, 4 can
be chosen arbitrarily due to a ``generalized gauge
invariance''.
By choosing the remaining two $n_i$ to correctly give
two of the partial amplitudes, non-trivial relations between
color-ordered amplitudes can be derived from the condition that the
remaining partial amplitudes are also reproduced correctly.  For
example,
\begin{equation}
s_{35} A_5^\tree(1,2,4,3,5) - (s_{13} + s_{23}) A_5^\tree(1,2,3,4,5)
           - s_{13} A_5^\tree(1,3,2,4,5)  = 0 \,.
\label{BCJAmplitudeRelation}
\end{equation}
This relation has generalizations for an arbitrary
number of external legs~\cite{BCJ}, which have been derived using string
theory~\cite{Bjerrum1}.

As discussed in refs.~\cite{Tye,Bjerrum2}, \eqn{BCJAmplitudeRelation}
is equivalent
to a relation that exhibits a Jacobi-like structure,
\begin{equation}
  {n_4 - n_1 + n_{15} \over s_{45}}
 - {n_{10} - n_{11} + n_{13} \over s_{24}}
 - {n_3 - n_1 + n_{12} \over s_{12}}
 - {n_5 - n_2 + n_{11} \over s_{51}} = 0 \,,
\label{GeneralizedJacobi}
\end{equation}
where only the sum over terms is required to vanish.  In this
equation, the Jacobi-like structure involves additional minus signs because
we follow the sign conventions given in ref.~\cite{BCJ} for the
expansion of the five-point amplitude,
\begin{eqnarray}
&&  A^\tree_5(1,2,3,4,5) \equiv {n_1\over s_{12}s_{45}}+{n_2\over s_{23}s_{51}}
+{n_3\over s_{34}s_{12}}+{n_4\over s_{45}s_{23}}+{n_5\over s_{51}s_{34}}\,, \nn\\
&&  A^\tree_5(1,2,4,3,5) \equiv {n_{12}\over s_{12}s_{35}}
+{n_{11}\over s_{24}s_{51}}-{n_3\over s_{43}s_{12}}
+{n_{13}\over s_{35}s_{24}}-{n_5\over s_{51}s_{43}}, \nn\\
&& A^\tree_5(1,3,2,4,5) = {n_{15} \over s_{13} s_{45}}
 - {n_{2} \over s_{23} s_{51}} - {n_{10} \over s_{24} s_{13}}
 - {n_{4} \over s_{45} s_{23}} - {n_{11} \over s_{51} s_{24}}\,.
\label{PartialAmplitudNumerator}
\end{eqnarray}
As explained in ref.~\cite{Tye,Bjerrum2}, relations of the
form~(\ref{GeneralizedJacobi}) are the natural gauge-invariant
numerator identities that emerge from string theory.  Because of the
generalized gauge invariance, these relations are less stringent than
the BCJ duality. Indeed, the individual terms in
\eqn{GeneralizedJacobi} are not required to vanish, but only their
sum.  We note that the heterotic string offers some important insight
into the BCJ duality (\ref{jacobin}): in the heterotic string both
color and kinematics arise from world-sheet fields, making the duality
more natural~\cite{Tye}.  Identities of the
form~(\ref{GeneralizedJacobi}), though interesting, will not play a
role in the analysis below.  In the remainder of this paper we will
only be concerned with numerators $n_i$ that satisfy the more
stringent BCJ-duality requirements of \eqn{jacobin}.

\subsection{Gravity squaring relations}

Another conjecture in ref.~\cite{BCJ} is that gravity tree amplitudes
can be constructed directly from the $n_i$ through ``squaring
relations''.  Consider two gauge-theory amplitudes,
\begin{eqnarray}
&& {1\over g^{n-2}}\,{\cal A}^\tree_n(1,2,3,\ldots,n)\,=\sum_{{\rm diags.}\, i}
                   {n_i c_i \over \prod_{\alpha_i} s_{\alpha_i}}\, , \nn \\
&& {1\over \tilde{g}^{n-2}}\, {\cal \widetilde{A}}^\tree_n(1,2,3,\ldots,n)\,=
         \sum_{{\rm diags.}\, i}
             {\n_i \tilde{c}_i \over \prod_{\alpha_i} s_{\alpha_i}} \, .
\label{GaugeTheoryLeftRight}
\end{eqnarray}
These two amplitudes do not have to be from the same theory, and can
have differing gauge groups and particle contents.  In ref.~\cite{BCJ}
the requirement that {\em both} the $n_i$ and the $\tilde{n}_i$ satisfy the
BCJ duality was imposed, \ie{} they satisfy all duality
 conditions
$n_i+n_j+n_k=0$ and $\tilde{n}_i+\tilde{n}_j+\tilde{n}_k=0$.  The
conjectured squaring relations state that gravity amplitudes are
given simply by
\begin{eqnarray}
&&  {-i \over (\kappa/2)^{n-2}}\, {\cal M}^\tree_n(1,2,3,\ldots,n)\,=
      \sum_{{\rm diags.}\,i} {n_i \n_i \over \prod_{j} s_{\alpha_i}} \, ,
\label{squaring}
\end{eqnarray}
where the sum runs over the same set of diagrams as in
\eqn{GaugeTheoryLeftRight}. The states appearing in the gravity theory
are just direct products of
gauge-theory states, and their
interactions are dictated by the product of the gauge-theory
momentum-space three-point vertices.
 The squaring relations~(\ref{squaring}) were
explicitly checked through
eight points and have recently been understood
from the KLT relations in heterotic string
theory~\cite{Tye}.

Using standard factorization arguments it is simple to see why one would
expect the BCJ duality to imply that gravity numerators are a
double copy of gauge-theory numerators.  Let us assume that the
numerators of all $n$-point gauge-theory amplitudes~(\ref{Anrep}) satisfy the
BCJ duality~(\ref{jacobin}).  Let us also assume that we have already proven that
the squaring relations~(\ref{squaring}) hold for
amplitudes with fewer legs.
Consider an ansatz for the $n$-point graviton amplitude given
in terms of diagrams by the double-copy formula~(\ref{squaring}). We now step
through
all possible factorization channels using real momenta.
By general field-theory considerations we know that in
each channel the diagrams break up into products of lower-point diagrams.
The sum over diagrams on each side of the factorization
pole forms a lower-point amplitude.   Since each numerator factor
of the $n$-point expression satisfies the duality condition,  we expect the lower-point tree
diagrams
on each side of the factorized propagator to inherit
this property when we choose
special kinematics to factorize a diagram.  Thus on each side of the pole we have a correct
set of double-copy numerators for the lower-point gravity amplitudes.
Stepping through all factorization channels we see  that
we have correct diagram-by-diagram factorizations in all channels.
This provides a strong indication that the double-copy property
follows from BCJ duality.
In \sect{SquaringSection}, we will make this conclusion rigorous using a BCFW construction.

\section{Invariances of Amplitudes and Generalized Squaring Relations}
\label{InvarianceSection}

In this section we discuss the invariances of gauge-theory and gravity
amplitudes. This leads to a new, more general squaring relation for
gravity, in which the numerators of only one of the gauge-theory
factors are required to satisfy the BCJ duality. (See also
ref.~\cite{BCJLoops}.)
As already noted, there is a substantial freedom in choosing the
numerators, which we will generically call generalized gauge invariance, even
though much of the freedom cannot be attributed to conventional gauge
invariance.  Our proof of the squaring relations will rely on an
understanding of the most general form of this freedom at $n$ points.

\subsection{Generalized gauge invariance}

Consider a shift of the $\tn_i$ in \eqn{GaugeTheoryLeftRight},
\begin{equation}
  \tn_i\to \tn_i+\Delta_i \,.
\label{GT}
\end{equation}
The key constraint that the $\Delta_i$ must satisfy is that they do
not alter the value of the amplitude, immediately leading to
\begin{equation}
\sum_{{\rm diags.}\ i}
{\Delta_i c_i \over \prod_{\alpha_i} s_{\alpha_i}} = 0\,.
\label{GT2}
\end{equation}
Any set of $\Delta_i$ that satisfies this constraint can be viewed as
a valid generalized gauge transformation since it leaves the amplitude
invariant.  Ordinary gauge transformations, of course, satisfy this
property.  We may take \eqn{GT2} as the fundamental constraint
satisfied by any generalized gauge transformation.

A key observation is that it is only the algebraic properties of the
$c_i$, and not their explicit values, that enter into the
cancellations in \eqn{GT2}.  This is so because the equation holds for
any gauge group. Thus any object that shares the algebraic properties
of the $c_i$ will satisfy a similar constraint.  Since the numerators
$n_i$ of the BCJ proposal satisfy exactly the same algebraic
properties as the $c_i$, we immediately have
\begin{equation}\label{identity}
 \sum_{\text{diags.}\ i}\,\frac{\Delta_i\,n_i}
  {\prod_{\alpha_i} s_{\alpha_i}}~=~0
\end{equation}
as the key statement of generalized gauge invariance.
This holds for any $\Delta_i$ that satisfies the constraint
(\ref{GT2}).  In particular, note that we do {\em not\,} need to require
the $\Delta_i$ to satisfy any Jacobi-like relations.

The freedom in making these shifts leads to an enormous freedom in
writing different representations of either gauge-theory or gravity
amplitudes.  In the gravity case, besides shifts of either the $n_i$
or the $\tn_i$, we can also shift the $n_i$ and $\tn_i$ simultaneously
as long as the interference terms vanish as well.

\subsection{A direct derivation of the identity}

It is instructive to directly demonstrate \eqn{identity} in a way that
goes beyond the explanation above.  If we take the $\Delta_i$ to be
local, then they move contributions between diagrams by canceling
propagators in such a way that they can be absorbed into other
diagrams.  We can thus decompose each $\Delta_i$ as
\begin{equation}\label{decDelta}
    \Delta_i=\sum_{\alpha_i}\Delta_{i,\alpha_i} s_{\alpha_i}\,,
\end{equation}
where the $\alpha_i$ label the different propagators in diagram $i$.
For simplicity, here we take the $\Delta_i$ to be local and linear in
inverse propagators, \ie to contain
no terms that are quadratic or higher order
in the inverse propagators $s_{\alpha_i}$. In this case, the
decomposition~(\ref{decDelta}) is unique because the inverse propagators of any diagram $i$ are independent under momentum conservation.

Consider three diagrams labeled by $i$, $j$, and $k$ whose color factors are
related by the Jacobi identity. These three diagrams
share all propagators except for one, as illustrated in
\fig{JacobiFigure}.  For definiteness, let us denote the distinct
inverse propagators of diagrams $i$, $j$, $k$ by $s$, $t$, and $u$,
respectively.
 Note that
$s+t+u\neq 0$ (except for four-point amplitudes).
 Instead, $s+t+u$ is the sum of the invariant
``masses'' of the four legs that enter the two vertices connected to the $s$
propagator in  diagram $i$ (which is the same as the four legs
entering the vertices of propagator $t$ in $j$, etc). If one of these
legs is external, its mass vanishes. Otherwise, this leg is another
internal propagator shared by the diagrams $i$, $j$, $k$ and its mass
simply the associated variable $s_\alpha$. Denoting the
invariant masses of  the four neighboring legs by $s_1$, $s_2$, $s_3$, and
$s_4$\,, we have
\begin{equation}\label{stu}
  s+t+u = s_1+s_2+s_3+s_4\,.
\end{equation}
Any color-ordered amplitude must contain either none or two of the
diagrams $i$, $j$, $k$. For definiteness, consider the color-ordered
amplitude that contains the diagrams $i$ and $j$. With the sign
conventions~(\ref{jacobin}), $n_i$ and $n_j$ must enter this
color-ordered amplitude with opposite sign.  The contributions of the
generalized gauge transformation~(\ref{GT}) to this color-ordered
amplitude is given by
\begin{equation}\label{deltaij}
 \frac{\Delta_{i,s}s}{\prod_{\alpha_i}s_{\alpha_i}} -
\frac{\Delta_{j,t}t}{\prod_{\alpha_j}s_{\alpha_j}}
 + \ldots\,
 ~=\, \frac{\Delta_{i,s}-\Delta_{j,t}}{\prod'_{\alpha_i}s_{\alpha_i}}
  + \ldots\,\,,
\end{equation}
where $\prod'_{\alpha_i}s_{\alpha_i}$ represents the product of
inverse propagators of diagram $i$ except for $s$,
i.e. ${\prod'_{\alpha_i}s_{\alpha_i}}=s^{-1}\prod_{\alpha_i}s_{\alpha_i}$. This
contribution must cancel by itself, because all other contributions,
 represented by  the `$\dots$' in \eqn{deltaij},
have a different propagator structure and are therefore independent
within this color-ordered amplitude. This independence is true because
the diagram $k$, which contributes to a different ordering, is
absent. (If we also had a contribution from diagram $k$, we could use
\eqn{stu} to relate contributions that have distinct propagator
structures.) We conclude that $\Delta_{i,s}=\Delta_{j,t}$. Repeating
this analysis for the other color-ordered amplitudes containing two of
the diagrams $i,j,k$, we obtain the constraints
\begin{equation}\label{delta}
    \Delta_{i,s}=\Delta_{j,t}\,,\qquad\Delta_{j,t}=\Delta_{k,u}\,, \qquad\Delta_{i,s}=\Delta_{k,u}
   \qquad\Longrightarrow\qquad\Delta_{i,s}=\Delta_{j,t}=\Delta_{k,u}\equiv\delta\,.
\end{equation}

Now we have assembled all the ingredients to prove~(\ref{identity}). With the decomposition~(\ref{decDelta}) for the $\Delta_i$,~(\ref{identity}) reads
\begin{equation}\label{newclaim}
    \sum_{\text{diags.}\, i} \frac{n_i\left( \sum_{\alpha_i}\Delta_{i,\alpha_i}s_{\alpha_i}\right)}{\prod_{\alpha_i}s_{\alpha_i}}=0\,.
\end{equation}
 Let us organize all terms in the decomposition~(\ref{newclaim}) according to their propagator structure.
For definiteness, we isolate the terms with the
inverse propagator structure ${\prod'_{\alpha_i}s_{\alpha_i}}$.
This gives
\begin{equation}
\begin{split}
   \frac{n_i\Delta_{i,s}+n_j\Delta_{j,t}+n_k\Delta_{k,u}}{\prod'_{\alpha_i}s_{\alpha_i}}
   =\delta\times \frac{n_i+n_j+n_k}{\prod'_{\alpha_i}s_{\alpha_i}}=0\,,
\end{split}
\label{cancelcontact}
\end{equation}
since we have taken the $n_i$ to satisfy the BCJ duality.  We can
repeat the same analysis for all other propagator structures appearing
in \eqn{newclaim}, and each of them vanishes separately. This then
explicitly exhibits the cancellation~(\ref{identity}) for local
$\Delta_i$ with linear contact terms.  For non-local $\Delta_i$, or
$\Delta_i$ with quadratic or higher contact terms, the cancellations
are similar but more involved.

\subsection{Generalized squaring relations}
We now apply the identity~(\ref{identity}) to find a generalization of
the BCJ squaring relations. The latter express the gravity amplitude
as
\begin{equation}
    {-i \over (\kappa/2)^{n-2}} \,{\cal M}_n~=\sum_{\text{diags.}\,i}\frac{n_i\tilde{n}_i}{\prod_{\alpha_i}  s_{\alpha_i}}\,,
\end{equation}
where both $n_i$ and $\tilde{n}_i$ are in the BCJ
representation. Consider now a set of gauge-theory numerators
$\tilde{n}'_i$ that do not satisfy the duality relations.  Defining
$\tilde{\Delta}_i=\tilde{n}'_i-\tilde{n}_i$, we find
\begin{equation}
\begin{split}
 {-i \over (\kappa/2)^{n-2}}\, {\cal M}_n&~=\sum_{\text{diags.}\,i}\frac{n_i\tilde{n}_i}{\prod_{\alpha_i}  s_{\alpha_i}}
   ~ =   \sum_{\text{diags.}\,i}\biggl[\,\frac{n_i\tilde{n}'_i}{\prod_{\alpha_i}  s_{\alpha_i}}~-~\frac{n_i\tilde{\Delta}_i}{\prod_{\alpha_i}  s_{\alpha_i}}\biggr]\,.
\end{split}
\end{equation}
It follows from the identity~(\ref{identity}) that the second term
vanishes. We thus conclude that
\begin{equation}\label{asym}
{-i \over (\kappa/2)^{n-2}}\, {\cal M}_n =  \sum_{\text{diags.}\,i}\frac{n_i\tilde{n}'_i}{\prod_{\alpha_i}  s_{\alpha_i}}\,,
\end{equation}
where the $ n_i$ satisfy the duality but the $\tilde{n}'_i$ do not
need to.  Interestingly, such asymmetric constructions work
just as well at loop level~\cite{BCJLoops}.

Note that we cannot also relax the BCJ duality condition on the $n_i$ in~(\ref{asym}). Indeed, performing an arbitrary
generalized gauge transformation $\Delta_i$ on the $n_i$ would create cross-terms of the form
\begin{equation}
   \sum_{\text{diags.}\,i}\frac{\Delta_i\tilde{n}'_i}{\prod_{\alpha_i}  s_{\alpha_i}}~\neq~0\,,
\end{equation}
which generically do not vanish because neither $\Delta_i$ nor  $\tilde{n}'_i$
satisfy the duality relations.


\section{Squaring relations between gauge and gravity theories}
\label{SquaringSection}

We now derive the squaring relations~(\ref{squaring}) between gravity
and gauge theories.  Our derivation requires two gauge theories whose
amplitudes have diagrammatic expansions with numerators that satisfy
the BCJ duality.  We show that the corresponding gravity numerators
are then simply the product of these gauge-theory numerators.  Our
proof relies on the existence of on-shell recursion relations for both
gravity and gauge theory based on the same shifted momenta. As we will
explain, this is for example satisfied for the pure Yang-Mills/gravity
pair in any dimension, and for the $\cn=4$ SYM/$\cn=8$ supergravity
pair in $D=4$.

\subsection{Derivation of squaring relations for tree amplitudes}
\label{secderivsq}

First, we consider the case of Einstein gravity obtained from two
copies of pure Yang-Mills theories.
The direct product of two Yang-Mills theories with $(D-2)$ states each (not counting color), gives $(D-2)^2$ states corresponding to a theory with a graviton, an anti-symmetric tensor and dilaton.  At tree level, however,
we can restrict ourselves to the pure-graviton sector since the other states do not enter as intermediate states.

We will assume that one can always obtain local Yang-Mills numerators
that satisfy the BCJ duality.
 We will prove inductively,
 using on-shell recursion relations with the lower-point amplitudes in the BCJ representation, that the $n$-point gravity numerator
 is the square of the $n$-point Yang-Mills numerator in the BCJ representation.

For three points, the squaring relations are trivial: there
is only one ``diagram'' with no propagators, and the relation simply
states~\cite{GSW}
\begin{equation}
    {-i \over \kappa/2}\,{\cal M}_3=   (A_3)^2\,,
\end{equation}
where $A_3$ is the color-ordered Yang-Mills 3-point amplitude.

For larger numbers of external legs, we proceed inductively.  To carry
out our derivation of the squaring relations we make use of on-shell
recursion relations.  These are derived using complex deformations of
the external momenta of the amplitude,
\begin{equation}\label{shift}
  p_a\to \hat p_a(z)=p_a + z q_a\, \qquad a=1,\ldots,n\,,\qquad  \hat p_a^2(z)=0\,,\qquad \sum_{a=1}^n q_a =0\,.
\end{equation}
Note that both momentum conservation and the on-shell conditions are
preserved.  To have valid recursion relations we demand that both the
gravity and the gauge-theory tree
amplitude vanish as we take the
deformation parameter to infinity:\footnote{Here, both gauge-theory factors are pure Yang-Mills amplitudes and thus $\tilde {\cal A}_n={\cal A}_n$. However, keeping the later generalization to other gravity/gauge-theory pairs in mind, we do not make use of this equality in the following discussion.}
  \begin{equation}\label{zfalloff}
    \hat {\cal M}_n(z)\to0\,,\qquad \hat {\cal A}_n(z)\to0\,,
    \qquad \hat {\tilde{\hskip -.15 cm{\cal A}}}_n(z)\to 0
    \qquad   \text{as } \quad z\to\infty\,.
  \end{equation}
The details of this complex shift (such as the number of shifted lines
or the particular choice of $q_a$) will not play a role in our
analysis, but we note that a large variety of shifts that satisfy~(\ref{zfalloff})
 are known~\cite{BCFW, Risager}.  The simplest of these are
BCFW two-line shifts.
At least one BCFW shift exists for any choice of two external lines
$a$ and $b$, such that both the gauge-theory and the gravity amplitude
vanish at large $z$~\cite{ArkaniHamed2008yf}.  Such shifts are also
known to work in $D\geq 4$ dimensions.  In our analysis we initially
pick one arbitrary (but fixed) such shift.

We also pick an arbitrary local choice of gauge-theory numerators
$n_i$ that satisfies the BCJ duality~(\ref{jacobin}).  The assumption
that such a choice exists at all is crucial for the following
derivation.  As the $n_i$ are local, their complex deformations $\hat
n_i(z)$ under the shift are polynomial in $z$; in particular $\hat
n_i(z)$ has no poles in $z$. (To ensure this property, one
  has to choose the polarization vectors such that they do not contain
  poles in $z$; such a choice is always possible.)

\begin{figure}[t]
\centerline{\epsfxsize 5 truein \epsfbox{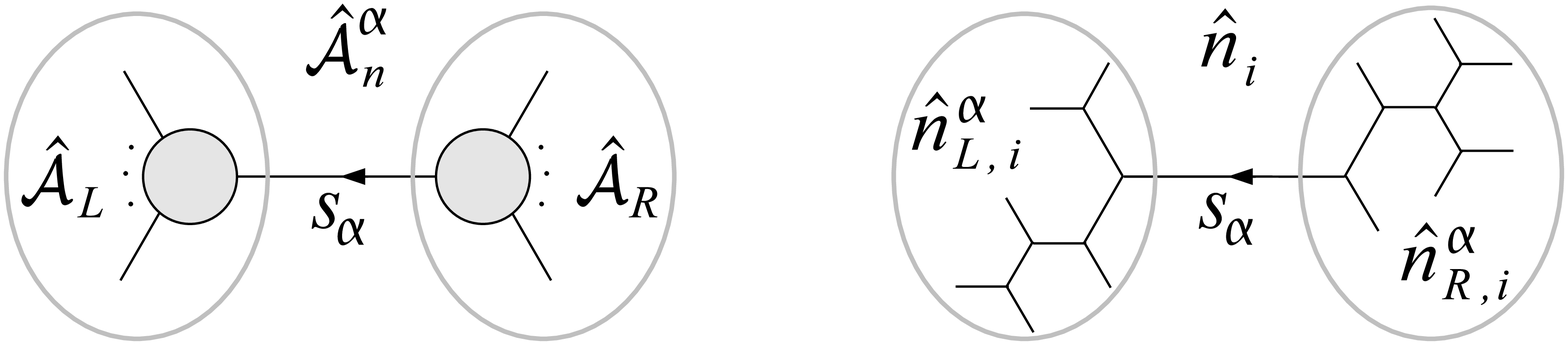}}
\centerline{(a)\hskip6.85cm (b)}
\caption[a]{\small (a) In an on-shell recursion relation, a given
  residue $\hat {\cal A}_n^\alpha$ is determined by diagrams sharing the same
  propagator labeled by $s_\alpha$.~~ (b) We can obtain a diagrammatic
  expansion of the recursion relation either from the numerators $\hat
  n_i$ of the shifted full amplitude $\hat {\cal A}_n$, or from the
  numerators $\hat n_{L,i}^\alpha$, $\hat n_{R,i}^\alpha$ of the
  subamplitudes.  }
\label{AlphaAmplFigure}
\end{figure}

As a first step, let us analyze the recursion relation for the
gauge-theory amplitude ${\cal A}_n$ that arises from the complex
shift. The amplitude contains poles at values $z=z_\alpha$ where an
internal propagator $1/s_\alpha$ goes on shell, \ie
$\hat{s}_\alpha(z_\alpha)=0$. We obtain an expression for ${\cal A}_n$
as a sum over residues,
\begin{equation}\label{Anresid}
    {\cal A}_n=\sum_{\alpha}\frac{\,\hat {\cal A}_n^\alpha\,}{s_\alpha}\,,\qquad \text{with } \hat {\cal A}_n^\alpha=i\,\hat {\cal A}_L(z_\alpha) \hat {\cal A}_R(z_\alpha)\,.
\end{equation}
This is illustrated in \fig{AlphaAmplFigure}a.  The residue
$\hat {\cal A}_n^\alpha$ at $z_\alpha$ with $\hat
s_\alpha(z_\alpha)=0$ factorizes into the product of a left and right
subamplitude, and it does not depend on the representation of the
amplitude. We can thus analyze each term in the sum over $\alpha$
separately, without ambiguity.  Note that in \eqn{Anresid} and from
now on, whenever there is a product of left/right factors, an implicit
sum over the polarizations of the on-shell intermediate state is
assumed.  Plugging in the left and right subamplitudes in the BCJ
representation, we obtain,
\begin{equation}\label{Analpha1}
        {1\over g^{n-2}}\,\hat {\cal A}_n^\alpha\,=\sum_{\alpha\dash{\rm diags.}\, i} \frac{i\,\hat n_{L,i}^\alpha\,\hat n_{R,i}^\alpha\,c_i}{\prod'_{\alpha_i}\hat s_{\alpha_i}(z_\alpha)}\,.
\end{equation}
Here, the sum goes only over diagrams $i$ that contain the channel
$\alpha$, and again the prime on the product indicates that the
propagator corresponding to that channel is not included,
${\prod'_{\alpha_i} s_{\alpha_i}}= {\prod_{\alpha_i\neq \alpha}
 s_{\alpha_i}}$.
The color factor $c_i$  arises from the color factors of the left and right subamplitudes after summing over the states of the intermediate gluon.
 The diagrammatic representation of the residue
$\hat {\cal A}_n^\alpha$ is illustrated in \fig{AlphaAmplFigure}b.

\begin{figure}[t]
\centerline{\epsfxsize 6 truein \epsfbox{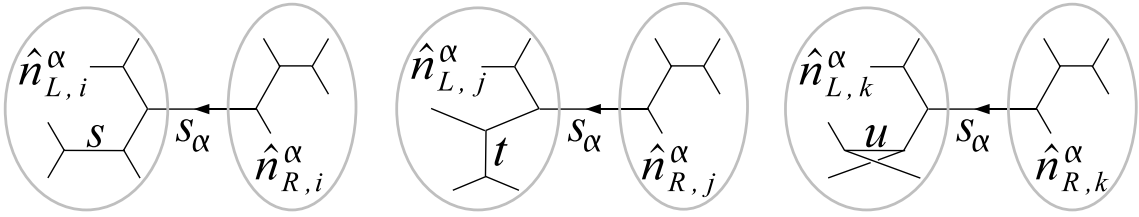}}
\caption[a]{\small The product $i\,\hat n_{L,i}\,\hat n_{R,i}$ satisfies the
duality relations satisfied by its factors $\hat n_{L,i}$ and $\hat n_{R,i}$.
}
\label{AlphaJacFigure}
\end{figure}

On the other hand, we can directly express $\hat {\cal A}_n^\alpha$ in terms of the original representation~(\ref{Anrep}) of ${\cal A}_n$. We obtain
\begin{equation}\label{Analpha2}
    {1\over g^{n-2}}\,\hat {\cal A}_n^\alpha\,=\sum_{\alpha\dash{\rm diags.}\, i}\frac{\hat n_{i}(z_\alpha)c_{i}}{\prod'_{\alpha_i}\hat s_{\alpha_i}(z_\alpha)}\,.
\end{equation}
Comparing \eqn{Analpha1} to \eqn{Analpha2}, we see that $i\,\hat
n_{L,i}^\alpha\,\hat n_{R,i}^\alpha$ and $\hat n_{i}(z_\alpha)$ are
related by some generalized gauge transformation:
\begin{equation}\label{compatible}
 \hat n_{i}(z_\alpha)=i\,\hat n_{L,i}^\alpha\,\hat n_{R,i}^\alpha+\Delta^\alpha_i\,,
\end{equation}
 where the $\Delta^\alpha_i$ satisfy
\begin{equation}
 \sum_{\alpha\dash{\rm diags.}\, i}
\frac{\Delta^\alpha_ic_{i}}{\prod'_{\alpha_i}\hat s_{\alpha_i}(z_\alpha)}~=~0\,.
\label{onshellGT}
\end{equation}
Note that the $\Delta^\alpha_i$ are only unambiguously defined at
$z=z_\alpha$, and should therefore not be thought of as a function of
$z$. Also note that this is not a single generalized gauge
transformation, but a distinct one for each choice of $\alpha$, and we
can analyze it separately for each $\alpha$. It will be important in
the following that the $\Delta^\alpha_i$ satisfy all duality
constraints that relate diagrams containing the internal line
$\alpha$, \ie
 \begin{equation}\label{jacobiD}
    \Delta^\alpha_i+\Delta^\alpha_j+\Delta^\alpha_k=0\,.
 \end{equation}
 To see this, note that $\Delta^\alpha_i =\hat n_{i}-i\,\hat
 n_{L,i}\,\hat n_{R,i}$\,, and as the $\hat n_i$ satisfy all duality
 relations it is sufficient to examine the duality properties of
 $i\,\hat n_{L,i}\,\hat n_{R,i}$.  The diagrams $i,j,k$
 in \eqn{jacobiD} share all but one propagator, and thus they either
 share the entire left or the entire right subdiagram of the
 factorized amplitude. For definiteness, let us consider the case
 where they share the entire right subdiagram, and thus $\hat
 n_{R,i}=\hat n_{R,j}=\hat n_{R,k}$. Then the duality
 relation~(\ref{jacobiD}) immediately follows from the corresponding
 duality relation for the numerators in the left subdiagram, $\hat
 n_{L,i}+\hat n_{L,j}+\hat n_{L,k}=0$\,. This is illustrated in
 \fig{AlphaJacFigure}.

 Note that we needed to introduce the $\Delta_i^\alpha$, which only
 satisfy a partial set of duality relations, because the recursion
 relation~(\ref{Analpha1}) by itself does not yield a BCJ-compatible
 representation of the amplitude.  In fact, from \eqn{Analpha1}, one
 can immediately read off an implied numerator representation $n_i'$
 of ${\cal A}_n$ given by
 \begin{equation}
     n'_i=\sum_{\alpha_i} i\,\hat n_{L,i}^{\alpha_i}\,\hat n_{R,i}^{\alpha_i}\prod_{\beta_i\neq\alpha_i}\frac{s_{\beta_i} }{ \hat s_{\beta_i}(z_{\alpha_i})}\,.
 \end{equation}
 Generically, these numerators $n'_i$ do not satisfy any duality relations.

We now turn to gravity.  Applying the recursion relation for gravity
to the amplitude ${\cal M}_n$, we obtain
\begin{equation}
    {\cal M}_n
    =\sum_\alpha\frac{i}{s_\alpha}\,\hat {\cal M}_L(z_\alpha) \hat {\cal M}_R(z_\alpha)\,.
\end{equation}
Using our inductive assumption that the squaring relations are valid
for lower-point amplitudes we can plug in the squaring
relations~(\ref{squaring}) for the subamplitudes $\hat {\cal M}_L$,
$\hat {\cal M}_R$ and obtain
\begin{equation}\label{Mn1}
    {1 \over (\kappa/2)^{n-2}}\, {\cal M}_n   =\sum_\alpha\frac{i}{s_\alpha}\sum_{\alpha\dash{\rm diags.}\, i}
    \frac{\bigl[i\,\hat n_{L,i}^\alpha\,\hat n_{R,i}^\alpha\,\bigr]\bigl[i\,\hat{\tilde{n}}_{L,i}^\alpha\,\hat{\tilde{n}}_{R,i}^\alpha\,\bigr]}
    {\prod'_{\alpha_{i}} \hat s_{\alpha_i}(z_\alpha)}\,.
\end{equation}
We can now use the gauge-theory relation~(\ref{compatible}) to rewrite
the gravity amplitude as
\begin{equation}\label{MnGT}
\begin{split}
    {1\over( \kappa/2)^{n-2}}\, {\cal M}_n
    &=\sum_\alpha\frac{i}{s_\alpha}\sum_{\alpha\dash{\rm diags.}\, i}
    \frac{\bigl[\hat n_{i}(z_\alpha)-\Delta^\alpha_i\bigr]\bigl[\hat{\tilde{n}}_{i}(z_\alpha)-\tilde{\Delta}^\alpha_i\bigr]}
    {\prod'_{\alpha_{i}} \hat s_{\alpha_{i}}(z_\alpha)}\\
    &=\sum_\alpha\frac{i}{s_\alpha} \sum_{\alpha\dash{\rm diags.}\, i}\Biggl[
   \frac{\hat n_{i}(z_\alpha)\hat{\tilde{n}}_{i}(z_\alpha)}{\prod'_{\alpha_{i}} \hat s_{\alpha_{i}}(z_\alpha)}
    ~-~\,\frac{\Delta^\alpha_i \,\hat{\tilde{n}}_{i}(z_\alpha)+\tilde{\Delta}^\alpha_i \,\hat{n}_{i}(z_\alpha)}{\prod'_{\alpha_{i}} \hat s_{\alpha_{i}}(z_\alpha)}
    ~+~\frac{\Delta^\alpha_i\tilde{\Delta}^\alpha_i}{\prod'_{\alpha_{i}} \hat s_{\alpha_{i}}(z_\alpha)}
    \Biggr]
    \,.
\end{split}
\end{equation}
The cross-terms involving the numerators $\tilde{\Delta}^\alpha_i\, \hat n_{i}(z_\alpha), \,\Delta^\alpha_i \,\hat{\tilde{n}}_{i}(z_\alpha)$ vanish due to the identity~(\ref{identity}), because the $n_i$ satisfy the BCJ duality. We will now argue that the last term also vanishes:
\begin{equation}\label{Delta2}
    \sum_{\alpha\dash{\rm diags.}\, i}\frac{\Delta^\alpha_i\tilde{\Delta}^\alpha_i}{\prod'_{\alpha_{i}} \hat s_{\alpha_{i}}(z_\alpha)}~=~0\,.
\end{equation}
To see this, we proceed analogously to the derivation of the
identity~(\ref{identity}), treating the factor $\Delta^\alpha_i$ as
the generalized gauge transformation, and the other factor
$\tilde{\Delta}^\alpha_i$ as the Jacobi-satisfying coefficient. We
have shown above that the $\Delta_i^\alpha$, $\tilde{\Delta}^\alpha_i$
satisfy the duality relations within the class of diagrams that contain
the line $\alpha$ (see eq~(\ref{jacobiD})). One may worry that this is
not sufficient to guarantee~(\ref{Delta2}), because there is also one
duality relation that relates the diagram $i$ to two diagrams in which
line $\alpha$ is replaced by its $t$- and $u$-channel analogue. These
diagrams do not contain the line $\alpha$ and thus do not appear
in \eqn{Delta2}. To see that this complication is harmless, we
expand $\Delta_i^\alpha$ in its distinct contact term contributions,
just like we expanded $\Delta_{i}$ in \eqn{decDelta}:
\begin{equation}
    \Delta_i^\alpha~=~ \sum_{\alpha_i\neq \alpha} \Delta_{i,\alpha_i}^\alpha \,\hat s_{\alpha_i} (z_\alpha)\,.
\end{equation}
Note that there is no contact term ambiguity in $\Delta_i^\alpha$
associated with $s_\alpha$ because $\hat s_\alpha(z_\alpha)=0$. In the
derivation of~(\ref{identity}), the duality relation between three
diagrams $i$, $j$, $k$ was important to cancel the contact term
ambiguities associated with the propagators $s$, $t$, $u$ (see
eq~(\ref{cancelcontact})). In our case there is no such ambiguity
associated with the propagator $s_\alpha$, so we only need the duality
relations~(\ref{jacobiD}) to argue that~(\ref{Delta2}) holds. We
conclude that,
\begin{equation}
    {-i\over( \kappa/2)^{n-2}}\, {\cal M}_n  =\sum_\alpha\frac{1}{s_\alpha}\sum_{\alpha\dash{\rm diags.}\, i}
    \frac{\hat n_{i}(z_\alpha)\, \hat \n_{i}(z_\alpha) }
    {\prod'_{ \alpha_{i}} \hat s_{\alpha_{i}}(z_\alpha)}\,.
\label{finalMn}
\end{equation}

We now define
\begin{equation}
 {-i \over (\kappa/2)^{n-2}}\,{\cal M}'_n~\equiv\sum_{{\rm diags.}\, i}\,\frac{n_i\,\tilde{n}_i}{\prod_{\alpha_i}  s_{\alpha_i}}\,,
\label{mz}
\end{equation}
where, as previously stated, the $n_i$ are the $n$-point Yang-Mills
numerator in the BCJ representation. Since the $n_i$ are local,
we see from~(\ref{finalMn}) that $\hat{\cal M}'_n(z)$ and $\hat
{\cal M}_n(z)$ have precisely the same pole structure.
Indeed, the falloff~(\ref{zfalloff}) of $\hat {\cal M}_n(z)$ at large $z$, together with~(\ref{finalMn}), imply that
\begin{equation}\label{Mnz}
  {-i \over (\kappa/2)^{n-2}}\, \hat{\cal M}_n (z) =\sum_\alpha\frac{1}{s_\alpha(z)}\sum_{\alpha\dash{\rm diags.}\, i}
  \frac{\hat n_{i}(z_\alpha) \,\hat \n_{i}(z_\alpha) }
  {\prod'_{ \alpha_{i}} \hat s_{ \alpha_{i}}(z_\alpha)}\,.
\end{equation}
 This form makes manifest that $\hat {\cal M}_n(z)$ and $\hat{\cal M}'_n(z)$ have coinciding residues for all finite-$z$ poles. However, in principle the two functions could
still differ by a function ${\cal P}$ of momenta and polarization
vectors that is polynomial in $z$ under the complex shift,
\begin{equation}
    {\cal P}~=~{\cal M}'_n- {\cal M}_n\,, \qquad \hat{\cal P}(z)~=~ \text{polynomial in $z$}\,\,.
    \label{PP}
\end{equation}
The on-shell gravity amplitude ${\cal M}_n$ is of course invariant under gravity on-shell gauge transformations, \ie under shifts of the polarization tensors that are proportional to the corresponding external graviton momentum. This gauge invariance actually also holds for ${\cal M}'_n$. To see this, note that we can break up the gravity gauge transformation into two Yang-Mills on-shell gauge transformations acting separately on the
 factors $n_i$ and $\tilde{n}_i$
in the numerators of ${\cal M}'_n$. We note that such a true on-shell gauge transformation leaves the duality property of the $n_i$ intact.\footnote{
To see this, recall that the $n_i$  are functions of momenta $p_a^\mu$ and polarization vectors $\epsilon_a^\mu$. A shift of the polarization vectors $\delta \epsilon_a^\mu\propto p_a^\mu$ treats all $n_i$ on equal footing, and thus can never spoil a duality relation $n_i+n_j+n_k=0$.}
 Invariance of ${\cal M}'_n$ then immediately follows from the identity~(\ref{identity}). We conclude that ${\cal P}$ is gauge-invariant.

The shift analysis above is not yet sufficient to argue that ${\cal P}$ is also local. While it cannot have poles in $z$, it could in principle have propagators in its denominator that are invariant under the particular shift that we chose. Note, however, that we could repeat the analysis above for any other shift under which both the gravity and the gauge-theory amplitude vanish. We conclude that ${\cal P}$ must be polynomial in $z$ under \emph{any} such shift. As explained above, many such shifts are available; in particular, there is a valid BCFW shift for any choice of two external lines. As no propagator can be invariant under all of these shifts, we conclude that ${\cal P}$ must be local.

From dimensional analysis, we also know that ${\cal P}$ must be quadratic in momenta. Note that this is only true because we are considering
 a gravity theory
that is not modified by higher-dimension operators. For example, if we had allowed for $\alpha'$ corrections to gravity, the expression  ${\cal P}$ could contain contributions that are higher order in momenta.

${\cal P}$ is thus a gauge-invariant, local expression quadratic in
momenta. No such expression exists because the matrix elements of any
contractions $D^mR^n$ of the Riemann tensor with covariant derivatives
contain at least $2n+m$ powers of momenta.\footnote{Another way to
  reach the same conclusion for the special case of $D=4$ dimensions
  is to note that a local, gauge-invariant expression must be
  expressible as a polynomial in the angle and square brackets of the
  spinor-helicity formalism. No expression quadratic in angle and
  square brackets can have the correct little-group scaling
  property~\cite{WittenTopologicalString} of an $n$-point amplitude.}
We conclude that
\begin{equation}
    {\cal P}~=~0\,,
\end{equation}
and therefore
\begin{equation}
    {-i \over (\kappa/2)^{n-2}}\, {\cal M}_n~=~ {-i \over (\kappa/2)^{n-2}}\,{\cal M}'_n~=\sum_{\text{diags.}\,i}\frac{n_i\tilde{n}_i}{\prod_{\alpha_i}  s_{\alpha_i}}\,,
\end{equation}
where both $n_i$ and $\tilde{n}_i$ are in the BCJ representation.
As discussed in section~\ref{InvarianceSection}, it then immediately follows from the identity~(\ref{identity}) that the squaring relations also hold if we only impose the duality relations on one of the two copies of gauge-theory numerators.  This concludes our derivation of the squaring relations for pure Einstein gravity in arbitrary dimensions $D\ge 4$.


\subsection{A five-point example}
We now illustrate
 some key
steps of our general derivation with the
simplest possible non-trivial example, the $D=4$\, five-point MHV
amplitude. At five points, a basis of color-ordered amplitudes
under the Kleiss-Kuijf relations~\cite{Kleiss} is given by
\begin{equation}
\begin{split}
  A_5(1,2,3,4,5)~&=~\frac{n_1}{s_{12}s_{45}}+\frac{n_2}{s_{23}s_{51}}+\frac{n_3}{s_{34}s_{12}}+\frac{n_4}{s_{45}s_{23}}+\frac{n_5}{s_{51}s_{34}}\,,\\
  A_5(1,4,3,2,5)~&=~\frac{n_6}{s_{14}s_{25}}+\frac{n_5}{s_{43}s_{51}}+\frac{n_7}{s_{32}s_{14}}+\frac{n_8}{s_{25}s_{43}}+\frac{n_2}{s_{51}s_{32}}\,,\\
  A_5(1,3,4,2,5)~&=~\frac{n_9}{s_{13}s_{25}}-\frac{n_5}{s_{34}s_{51}}+\frac{n_{10}}{s_{42}s_{13}}-\frac{n_8}{s_{25}s_{34}}+\frac{n_{11}}{s_{51}s_{42}}\,,\\
  A_5(1,2,4,3,5)~&=~\frac{n_{12}}{s_{12}s_{35}}+\frac{n_{11}}{s_{24}s_{51}}-\frac{n_{3}}{s_{43}s_{12}}+\frac{n_{13}}{s_{35}s_{24}}-\frac{n_{5}}{s_{51}s_{43}}\,,\\
  A_5(1,4,2,3,5)~&=~\frac{n_{14}}{s_{14}s_{35}}-\frac{n_{11}}{s_{42}s_{51}}-\frac{n_7}{s_{23}s_{14}}-\frac{n_{13}}{s_{35}s_{42}}-\frac{n_2}{s_{51}s_{23}}\,,\\
  A_5(1,3,2,4,5)~&=~\frac{n_{15}}{s_{13}s_{45}}-\frac{n_{2}}{s_{32}s_{51}}-\frac{n_{10}}{s_{24}s_{13}}-\frac{n_{4}}{s_{45}s_{32}}-\frac{n_{11}}{s_{51}s_{24}}\,,\\
\end{split}
\end{equation}
where we follow the notation of ref.~\cite{BCJ}, including the signs
in the duality relations.  Specifying the
negative-helicity lines to be $1$ and $5$, we can compute the first
two color-ordered amplitudes above directly from the Parke-Taylor~\cite{PT}
formula:
\begin{eqnarray}
&& A_5(1^-2^+3^+4^+5^-)= i \frac{\<15\>^4}{\<12\>\<23\>\<34\>\<45\>\<51\>}\,,
   \nn \\
&& A_5(1^-4^+3^+2^+5^-)=i \frac{\<15\>^4}{\<14\>\<43\>\<32\>\<25\>\<51\>}\,.
\end{eqnarray}
Let us furthermore specify the BCFW shift $[1,5\>$ for the following analysis,
\begin{equation}\label{BCFW}
    |1]~\to~|\hat 1]=|1]-z|5]\,, \qquad |5\>~\to~|\hat 5\>=|5\>+z|1\>\,.
\end{equation}
We could start with a general local expression for the $n_i$ to ensure that they are polynomial in $z$ under the shift. In fact, there is a $225$-parameter family of such local $n_i$ that satisfy the duality relations and reproduce all Kleiss-Kuijf relations~\cite{Kleiss} correctly. For our purposes, however, it is more convenient to construct a simple choice of $n_i$ by hand. To reproduce the first amplitude $A_5(1^-2^+3^+4^+5^-)$ correctly, we  pick
\begin{equation}
    n_1=- i\frac{\<15\>^3}{\<23\>\<34\>}\times [12][45]\,,\qquad n_2=n_3=n_4=n_5=0\,.
\end{equation}
It is obvious that $n_1$ is polynomial under the BCFW shift~(\ref{BCFW}), because this shift leaves the angle brackets $\<23\>$ and $\<34\>$  invariant. The duality relations
$n_4-n_2+n_7=0$ and $n_3-n_5+n_8=0$ immediately imply that $n_7$ and $n_8$ vanish. To reproduce the second amplitude $A_5(1^-4^+3^+2^+5^-)$ correctly, we thus need to  set
\begin{equation}
\begin{split}
  n_6=- i\frac{\<15\>^3}{\<23\>\<34\>}\times [14][25]\,,\qquad n_7=n_8=0\,.
\end{split}
\end{equation}
$n_6$ is also manifestly polynomial under the specified BCFW
shift. All other numerators $n_9,\ldots,n_{15}$ are now determined
through the duality relations and are of course also polynomial under
the shift. In summary, we obtain the following choice of numerators:
\begin{equation}
\begin{split}
  n_1&=n_{12}=n_{15}=- i\frac{\<15\>^3}{\<23\>\<34\>}\times[12][45]
   \,,\qquad \qquad~ n_6=n_9=n_{14}=-
   i\frac{\<15\>^3}{\<23\>\<34\>}\times[14][25]\,,\\
 n_{10}&=-n_{13}=
   i\frac{\<15\>^3}{\<23\>\<34\>}\times\bigl([12][45]-[14][25]\bigr)\,,
   \qquad n_2=n_3=n_4=n_5=n_7=n_8=n_{11}=0\,.
\end{split}
\end{equation}
Let us now consider the contribution of the factorization channel $s_\alpha=s_{45}$ to the BCFW recursion relation. All shifted expressions must thus be evaluated at $z=z_\alpha \equiv -[45]/[14]$\,.
The amplitude factorizes into a ``right'' three-point anti-MHV subamplitude\footnote{We adopt the convention that all external momenta are incoming, and the internal momentum $P$ is incoming in the left subamplitude, and outgoing in the right subamplitude. We use the spinor conventions $|\text{-}P\>=i|P\>$, $|\text{-}P]=i|P]$, $s_{ab}=-[ab]\<ab\>$.}
\begin{equation}
 \hat A_R( 4^+,\hat 5^-, -\hat P^+)~=~\hat n_R~=~- i \frac{[4,-\hat P]^4}{[45][5,-\hat P][-\hat P, 4]}\,,
\end{equation}
and a ``left'' four-point subamplitude. The latter is MHV and takes the form
\begin{equation}
    \hat A_L(\hat 1^-,2^+,3^+,\hat P^-)=\frac{\hat n_{L,s}}{\hat s_{ 12}}-\frac{\hat n_{L,t}}{\hat s_{23}}\,.
\end{equation}
We can pick an arbitrary (local or non-local) representation of $\hat
n_{L,s}$,  $\hat n_{L,t}$.
One choice is given by,
\begin{equation}
    \hat n_{L,s}= i\frac{\<12\>\<1\hat P\>[23]^2}{2\,\<23\>[3\hat P]}\,,\qquad \hat n_{L,t}= -i\frac{\<1\hat P\>[ 23]^3}{2\, [\hat 12][3\hat P]}\,,
    \qquad \hat n_{L,u}=- \hat n_{L,s}-\hat n_{L,t}\,.
\end{equation}
Combining left and right subamplitudes, the color factors of $\hat n_{L,s} \hat n_R$, $\hat n_{L,t} \hat n_R$, and $\hat n_{L,u} \hat n_R$, are $c_1$, $c_4$, and $c_{15}$, respectively. Their corresponding numerators satisfy the duality relation $n_{1}-n_{4}-n_{15}=0$. We thus define
\begin{equation}
\begin{split}
    \Delta_{1}^\alpha\,=\hat n_1\,-i\,\hat n_{L,s}\hat n_R\,
    &=- i\frac{\<1\hat 5\>^3[\hat 12][45]}{\<23\>\<34\>} - i\frac{\<12\>\<1\hat P\>[4\hat P]^3[23]^2}{2\,\<23\>[3\hat P][45][5\hat P]}
    = i\frac{\<1\hat 5\>^3[45]}{2\,\<12\>\<23\>\<34\>}\times \hat s_{1 2}\,,\\[1ex]
    \Delta_{4}^\alpha\,=\hat n_4\,+i\,\hat n_{L,t}\hat n_R\,
    &=-i\frac{\<1\hat P\>[\hat 23]^3[4\hat P]^3}{2\, [\hat 12][3\hat P][45][5\hat P]}
    = -i\frac{\<1\hat 5\>^3[45]}{2\, \<12\>\<23\>\<34\>} \times \hat s_{23}\,,   \\[1ex]
    \Delta_{15}^\alpha=\hat n_{15}+i\,\hat n_{L,u}\hat n_R\,
    &=-i\frac{\<1\hat 5\>^3[45]}{2\, \<12\>\<23\>\<34\>} \times \hat s_{13} \,.
\end{split}
\end{equation}
Note that these $\Delta_i^\alpha$ indeed satisfy the duality relation
on the pole: $\Delta_{1}^\alpha-\Delta_4^\alpha-\Delta_{15}^\alpha=0$.

The crucial step in our derivation of the squaring relations was the
cancellation of the $\Delta_i^\alpha\hat n_i$ and
$(\Delta_i^\alpha)^2$ pieces in \eqn{MnGT}. It is now
straightforward to verify this cancellation directly in our current
example:
\begin{equation}
\begin{split}
  \frac{\Delta_{1}^\alpha\hat n_1}{\hat s_{1 2}}+\frac{\Delta_{4}^\alpha\hat n_4}{\hat s_{2 3}}+\frac{\Delta_{15}^\alpha\hat n_{15}}{\hat s_{1 3}}
  &= i \, \frac{\<1\hat 5\>^3[45]}{2\,\<12\>\<23\>\<34\>}\times\bigl(\hat n_1-\hat n_4-\hat n_{15}\bigr)=0\,,\\[1.5ex]
  \frac{(\Delta_{1}^\alpha)^2}{\hat s_{1 2}}+\frac{(\Delta_{1}^\alpha)^2}{\hat s_{2 3}}+\frac{(\Delta_{15}^\alpha)^2}{\hat s_{13}}
  &=-\Biggl(\frac{\<1\hat 5\>^3[45]}{2\,\<12\>\<23\>\<34\>}\Biggr)^2\times\bigl(\hat s_{1 2}+\hat s_{23}+\hat s_{13}\bigr)=0\,,
\end{split}
\end{equation}
where we used the pole condition $\hat s_{1 23}=\hat s_{45}=0$.

\subsection{Generalization to other gravity/gauge-theory pairs}

The derivation of the squaring relations in the previous section specifically pertained to pure gravity and pure Yang-Mills theory. However, only a few steps in the  derivation depended on this specific choice of theories. For a more general gravity/gauge-theory pair, the above derivation goes through if the following three conditions are satisfied:
\begin{enumerate}
  \item Every amplitude in the gauge theory can be expressed using local numerators that satisfy the BCJ duality.
  \item There exist ``valid'' complex shifts of the external momenta, \ie shifts such that both gauge-theory and gravity tree amplitudes vanish at large $z$. Such shifts give rise to on-shell recursion relations.
  \item Each propagator of every gravity amplitude must develop a pole under at least one of these valid complex shifts. This property was crucial for our conclusion above that ${\cal P}$ defined in \eqn{PP} vanishes identically.
\end{enumerate}

An interesting candidate gravity/gauge-theory pair are the $\cn=4$ SYM
and $\cn=8$ supergravity theories in four dimensions. Just as for pure
Yang-Mills theory, it remains to be shown that amplitudes in $\cn=4$
SYM can be expressed using local numerators satisfying the BCJ
duality. Although we expect the duality to work in supersymmetric
theories~\cite{BCJ,Mafra,Sondergaard}, naively, the conditions (2) and (3) above
seem hard to satisfy; while each $\cn=4$ SYM amplitude with $n>4$
external legs admits at least one valid BCFW
shift~\cite{Cheung,Elvang} and a variety of valid holomorphic
shifts~\cite{Elvang,Elvang2,Kiermaier2009yu}, the same does not hold
for the amplitudes of $\cn=8$ supergravity~\cite{Bianchi2008pu}. For
certain amplitudes, we seem to have no valid BCFW shifts available at
all, let alone sufficiently many to conclude that ${\cal P}$ vanishes.

Fortunately, there is a simple fix to this problem: We promote the
numerators $n_i$ to on-shell superfields ${\mathfrak n}_i$ and the
amplitudes ${\cal A}_n$, ${\cal M}_n$ to superamplitudes ${\mathfrak
  A}_n$, ${\mathfrak M}_n$\,, which depend on Grassmann parameters
$\eta_{a,A}$ (where $a$ and $A$ denote the particle index and the
$SU(\cn)$ index, respectively).  The superamplitudes ${\mathfrak A}_n$
and ${\mathfrak M}_n$ are $\eta$-polynomials that encode all $n$-point
amplitudes of SYM and supergravity as their coefficients.

At the MHV level, we can circumvent a new derivation of the squaring
relations altogether. The tree-level pure-gluon amplitudes of SYM are
identical to the ones of pure Yang-Mills theory. The pure-graviton
amplitudes in supergravity and pure gravity also coincide. The
squaring relations then immediately apply, in particular, to the
gluon/graviton MHV amplitude pair ${\cal A}_n^{--+\cdots+}$, ${\cal
  M}_n^{--+\cdots+}$. Choosing duality-satisfying numerators
$n_i^{--+\cdots+}$ for the gluon amplitude, we define
``super-numerators''
\begin{equation}
    {\mathfrak n}_i~=~\frac{\delta^{(8)}\bigl(\bar Q_A\bigr)}{\<12\>^{8}}\times n_i^{--+\cdots+}\,, \qquad \bar Q_A=\sum_a\,|a\>\eta_{a,A}\,.
\end{equation}
These super-numerators satisfy all duality relations, because
\begin{equation}
    {\mathfrak n}_i+{\mathfrak n}_j+{\mathfrak n}_k~\propto~ n_i^{--+\cdots+}+n_j^{--+\cdots+}+n_k^{--+\cdots+}~=~0\,.
\end{equation}
The ${\mathfrak n}_i$, though non-local, also manifestly satisfy the squaring relations:
\begin{equation}
\begin{split}
  {-i \over \kappa^{n-2}}\, {\mathfrak M}_n&~=~ \frac{-i\delta^{(16)}\bigl(\bar Q_A\bigr)}{\kappa^{n-2}\,\<12\>^{16}}\,{\cal M}_n^{--+\cdots+}
  ~=~\frac{\delta^{(16)}\bigl(\bar Q_A\bigr)}{\<12\>^{16}}\times \sum_i\frac{(n_i^{--+\cdots+})^2}{\prod_{\alpha_i}s_{\alpha_i}}
  ~=~\sum_i\frac{{\mathfrak n}_i \tilde{{\mathfrak n}}_i}{\prod_{\alpha_i}s_{\alpha_i}}\,.
\end{split}
\end{equation}
It then  follows from the identity~(\ref{identity}) that all BCJ numerators must satisfy the squaring relations at the MHV level.

 Beyond the MHV level we make use of~\cite{ArkaniHamed2008gz,Brandhuber2008pf}, where it was shown that the  superamplitudes ${\mathfrak A}_n$ and ${\mathfrak M}_n$ vanish under a {\em super}-BCFW shift of any two lines $a$ and $b$:
\begin{equation}
        |a]~\to~|\hat a]=|a]-z|b]\,, \qquad |b\>~\to~|\hat b\>=|b\>+z|a\>\,, \qquad \eta_{a,A}\to\hat \eta_{a,A}=\eta_{a,A}-z\,\eta_{b,A}\,.
\end{equation}
We thus have a large number of valid super-BCFW shifts available for
the superamplitudes ${\mathfrak A}_n$ and ${\mathfrak M}_n$, and conditions (2)
and (3) are easily satisfied for this gravity/gauge-theory pair.
Instead of performing sums over intermediate polarizations in the
derivation of section~\ref{secderivsq} (for example in the product
$\hat n_{L,i}\hat n_{R,i}$ of \eqn{Analpha1}), we perform
integrals over the Grassmann parameters $\eta_{P,A}$ associated with
the internal line:
\begin{equation}
 \hat n_{L,i}^\alpha\,\hat n_{R,i}^\alpha\quad\to\quad
 \int d^4\eta_{P,A}\,\,\hat {\mathfrak n}_{L,i}^\alpha\,\hat {\mathfrak n}_{R,i}^\alpha \,.
\end{equation}
The remaining analysis carries through without modification,
establishing the squaring relations for $\cn=4$ SYM/$\cn=8$
supergravity.

A similar analysis can be repeated for other gravity/gauge-theory
pairs by systematically verifying the conditions (1)--(3)
above. Whether the KLT relations are valid for a particular
gravity/gauge-theory pair is usually addressed using the $\alpha'\to
0$ limit of string theory amplitudes.  Our three conditions above for
the squaring relations, on the other hand, give purely field-theoretic
criteria for the validity of ``gravity = (gauge theory)$\times$(gauge
theory)''.

\subsection{Extension to loops}

We note that our tree-level derivation of the squaring relations (\ref{squaring}) from the  BCJ duality (\ref{jacobin})
 immediately extends  to loops via the unitarity method~\cite{UnitarityMethod}.
   In the unitarity method, no
shifts of momenta are required and there is no issue with large-$z$
behavior, if the cuts are evaluated in $D$ dimensions, ensuring cut
constructability~\cite{DdimUnitarity}.  Assuming that gauge-theory
loop amplitudes satisfy the duality, a gravity ansatz in
terms of diagrams built by taking double copies of numerators will
have all the correct cuts in all channels, since the numerators of all
tree diagrams appearing in the cuts are double copies.  This
immediately implies that the gravity amplitude so constructed is
correct.

Given two gauge theories whose $L$-loop numerators $n_i,\,\tilde{n}_i$
can be arranged to satisfy the BCJ duality, and whose tree amplitudes
are related through the squaring relations to a corresponding gravity
theory, we can immediately write down the gravity $L$-loop
amplitude~\cite{BCJLoops}:
\begin{equation}
  {(-i)^{L+1} \over (\kappa/2)^{n+2L-2}}{\cal M}_n^\text{$L$-loop}~=
\sum_{\text{diags.}\, i}\,\int\prod_{a=1}^L \frac{d^Dl_a}{(2\pi)^D}\,\,
 {n_i\,\tilde{n}_i \over \prod_{\alpha_i} s_{\alpha_i}}\,,
\end{equation}
where the numerators $n_i\,\tilde{n}_i$ and propagators
$1/s_{\alpha_i}$ depend on external and loop momenta, and the sum runs
over all $L$-loop diagrams with only three-point vertices.  We note
that this squaring relation holds at loop level for arbitrary loop
momenta, while the traditional KLT relations only apply to unitarity
cuts that factorize the loop amplitude into a product of tree
amplitudes.

In the next section we construct Lagrangians
whose diagrams reflect the BCJ duality, suggesting that the
gauge-theory duality does indeed extend to loop level. Interestingly,
not only has the extension of the duality to loop level been
explicitly demonstrated in a pure Yang-Mills two-loop example and an
$\NeqFour$ super-Yang-Mills three-loop example, but the double-copy
property of the corresponding gravity loop amplitudes has also been
confirmed~\cite{BCJLoops}.

\section{A Lagrangian generating diagrams with BCJ duality}
\label{LagrangianSection}

We now turn to the question of finding a Lagrangian which generates
amplitudes with numerators that manifestly satisfy the BCJ duality.
If a local Lagrangian of this type could be found, it would enable us
to construct a corresponding gravity Lagrangian whose squaring
relations with Yang-Mills theory are manifest.  We show that such a
construction is indeed possible, and we present the explicit form of a
Yang-Mills Lagrangian which generates diagrams that respect the BCJ
duality up to five points. We use it to construct the corresponding
Lagrangian for gravity. We also outline the structure of Lagrangians
that preserve the duality in higher-point diagrams.

\subsection{General strategy of the construction}

A Yang-Mills Lagrangian with manifest BCJ duality can only differ from
the conventional Yang-Mills Lagrangian by terms that do not affect
the amplitudes.  The amplitudes are unaffected, for example, by adding
total derivative terms or by carrying out field redefinitions.  In
fact, the MHV Lagrangian~\cite{Gorsky} for the CSW~\cite{CSW}
expansion is an example where identities or structures of tree-level
amplitudes can be derived through a field redefinition of the original
Lagrangian.  Such a construction has the additional complication that
a Jacobian can appear at loop level.  Surprisingly, we find that not
only does a Lagrangian with manifest BCJ duality exist, it differs
from the conventional Lagrangian by terms whose sum is identically
zero by the color Jacobi identity!  Although the sum over added terms
vanishes, they cause the necessary rearrangements so that the BCJ
duality holds.  Another curious property is that the additional terms
are necessarily non-local, at least if we want a covariant Lagrangian
without auxiliary fields.

For example, consider five-gluon tree amplitudes. To obtain diagrams
that satisfy the BCJ duality one is required to add terms to the
Lagrangian of the form
\begin{equation}
\mathcal{L}'_5\sim \Tr\,[A^{\nu} ,A^{\rho}]
\frac{1}{\square}\bigl([[\partial_{\mu} A_{\nu}, A_{\rho}], A^{\mu}]
+[[A_{\rho}, A^{\mu}],\partial_{\mu} A_{\nu}]
+[[A^{\mu},\partial_{\mu} A_{\nu} ], A_{\rho}]\bigr),
\end{equation}
along with other contractions. If we expand the commutators,
the added terms immediately vanish
by the color Jacobi identity.
If, however, the commutators are re-expressed in terms
of group-theory structure constants, they generate terms that
get distributed
across different diagrams and color factors.
We find similar results up to six points, suggesting that it
is a general feature for any number of points.

Since the BCJ duality relates the structure of kinematic numerators
and color factors of diagrams with only three-point vertices, the
desired Lagrangian should contain only three-point interactions.  To
achieve this we introduce auxiliary fields.  The auxiliary fields not
only reduce the interactions down to only three points, they also
convert the newly introduced non-local terms into local interactions.
This procedure introduces a large set of auxiliary fields into the
Lagrangian.  This is not surprising since we want a double copy of
this Lagrangian to describe gravity. The ordinary gravity Lagrangian
contains an infinite set of contact terms; if we were to write it in
terms of three-point interactions we would need to introduce a new set
of auxiliary fields for each new contact term in the expansion.  Since
in our approach the gravity Lagrangian is simply the square of the
Yang-Mills Lagrangian, it is natural to expect that the desired
Lagrangian contains a large (perhaps infinite) number of auxiliary
fields.  We now begin our construction of a Yang-Mills Lagrangian with
manifest BCJ duality.

\subsection{The Yang-Mills Lagrangian through five points}

We write the Yang-Mills Lagrangian as
\begin{equation}
  \mathcal{L}_{YM} = \mathcal{L}+\mathcal{L}'_5+ \mathcal{L}'_6+\ldots
\end{equation}
where $\mathcal{L}$ is the conventional Yang-Mills Lagrangian and
$\mathcal{L}'_n,\,n>4$ are the additional terms required to satisfy
the BCJ duality. At four points, the BCJ duality is trivially
  satisfied in any gauge~\cite{BCJ}, so $\mathcal{L}$ by itself
  will generate diagrams whose numerators satisfy \eqn{jacobin}.
  For simplicity we choose
  Feynman gauge for $\mathcal{L}$,\footnote{We are considering only tree
  level at this point. Therefore we ignore ghost terms.} though
similar conclusions hold for other gauges. All contact terms are
uniquely assigned to the three-vertex diagram carrying the
corresponding color factor.  The $\mathcal{L}'_n$ are required to leave
scattering amplitudes invariant, and they must rearrange the numerators
of diagrams in a way so that the BCJ duality is satisfied.  It turns out
that the set of terms with the desired properties is not unique.
Indeed, ``self-BCJ'' terms that satisfy the BCJ duality by themselves
can also be added.  This ambiguity is related to the residual ``gauge
invariance'' that remains after solving the duality identities
\cite{BCJ,Tye}.

By imposing the constraint that the generated five-point diagrams
satisfy the BCJ duality~(\ref{jacobin}), we find the following Lagrangian:
\begin{align}
  \mathcal{L} &= \tfrac{1}{2} A^a_\mu \square A^{a\mu} -g f^{a_1a_2a_3} \partial_{\mu}A^{a_1}_{\nu} A^{a_2\mu} A^{a_3\nu}-\tfrac{1}{4} g^2 f^{a_1a_2b}f^{ba_3a_4} A^{a_1}_\mu A^{a_2}_\nu A^{a_3\mu} A^{a_4\nu}\nonumber \\
  \mathcal{L}'_5
     &=- \tfrac{1}{2} g^3 f^{a_1a_2b}f^{ba_3c}f^{ca_4a_5}  \nonumber \\
	&\quad \times \left( \partial^{\phantom{a_1}}_{[\mu} A^{a_1}_{\nu]} A^{a_2}_\rho A^{a_3\mu} +
	\partial^{\phantom{a_2}}_{[\mu} A^{a_2}_{\nu]} A^{a_3}_\rho A^{a_1\mu} +
	\partial^{\phantom{a_3}}_{[\mu} A^{a_3}_{\nu]} A^{a_1}_\rho A^{a_2\mu} \right)
	\frac{1}{\square}(A^{a_4\nu} A^{a_5\rho}) \,.
\end{align}
The numerators $n_i$ are derived from this action by first computing
the contribution from the three-point vertices, which gives a set of
three-vertex diagrams with unique numerators. Then the contributions
from the four- and five-point interaction terms are assigned to the
various diagrams
 with only three-point vertices
according to their color factors. Since these terms
will contain fewer propagators than those obtained by using only
three-point vertices, their contributions to the numerators contain
inverse propagators.
 Finally, we combine all diagrams with the same color factor, however they arose in the procedure above, into a single diagram. Its kinematic coefficient is the desired numerator that satisfies the BCJ duality.
In this light, the purpose of $\mathcal{L}'_5$ is
to restore the BCJ duality (\ref{jacobin}) violated by the interaction
terms of $\mathcal{L}$.

Although $\mathcal{L}'_5$ is not explicitly local, as we mentioned, we
can make it local by the introduction of auxiliary fields.  It turns
out that without auxiliary fields there is no solution for a local
Lagrangian in any covariant gauge that generates numerators satisfying
the BCJ duality.  The non-locality explains the difficulty of
stumbling onto this Lagrangian without knowing its
desired property ahead of time.

As previously mentioned, $\mathcal{L}'_5$ is identically zero by the color
Jacobi identity. To see this we can relabel color indices to obtain
\begin{align}
 \mathcal{L}'_5 &= -\tfrac{1}{2} g^3 (f^{a_1a_2b}f^{ba_3c} + f^{a_2a_3b}f^{ba_1c}
   + f^{a_3a_1b}f^{ba_2c})f^{ca_4a_5} \nonumber \\&\quad
   \times \partial^{\phantom{a_1}}_{[\mu} A^{a_1}_{\nu]} A^{a_2}_\rho A^{a_3\mu}
  \frac{1}{\square}(A^{a_4\nu} A^{a_5\rho}) \,.
\label{Lagragian5Alt}
\end{align}
As apparent in (\ref{Lagragian5Alt}), the canceling terms have
different color factors and thus appear in different channels. For the
individual diagrams these terms are non-vanishing. Furthermore, they
alter the numerators of the individual diagrams such that the BCJ
duality (\ref{jacobin}) is satisfied.

It is interesting to note that there is one other term that can be
added to Yang-Mills at five points which preserves the relation
(\ref{jacobin}):
\begin{align}
    \mathcal{D}_5
     &= \tfrac{-\beta}{2} g^3 f^{a_1a_2b}f^{ba_3c}f^{ca_4a_5}\nn \\
 &\quad \times \left( \partial^{\phantom{a_1}}_{(\mu} A^{a_1}_{\nu)} A^{a_2}_\rho A^{a_3\mu} +
\partial^{\phantom{a_2}}_{(\mu} A^{a_2}_{\nu)} A^{a_3}_\rho A^{a_1\mu} +
\partial^{\phantom{a_3}}_{(\mu} A^{a_3}_{\nu)} A^{a_1}_\rho A^{a_2\mu} \right)
\frac{1}{\square}(A^{a_4\nu} A^{a_5\rho}) \,,
\label{DFive}
\end{align}
where $\beta$ is an arbitrary parameter. $\mathcal{D}_5$ also vanishes
identically by the color Jacobi identity.  Since $\mathcal{D}_5$
does not serve to correct lower-point contributions to make the BCJ
duality relations hold through five points, we do not need to include
it. It does however show that there are families of Lagrangians with
the desired properties.


\subsection{Towards gravity}

Now that we have a Lagrangian that gives the desired numerators $n_i$
for gauge theory (\ref{GaugeTheoryLeftRight}), we use it to construct
the tree-level gravity Lagrangian by demanding that it gives diagrams
whose numerators are a double copy of the gauge-theory numerators, as
in \eqn{squaring}.  However, as explained above, we need to first
bring the Yang-Mills Lagrangian into a cubic form to achieve this.  We
can do so by introducing an auxiliary field $B_{\mu\nu}^a$:
\begin{equation}
 \mathcal{L}_{YM} = \tfrac{1}{2} A^{a\mu}\square A_{\mu}^{a}
  +B^{a\mu\nu} B_{\mu\nu}^{a}
  - g f^{abc}(\partial_{\mu} A_{\nu}^{a}
  + B_{\mu\nu}^{a}) A^{b\mu} A^{c\nu}\,.
\end{equation}
This is equivalent to the ordinary Yang-Mills Lagrangian as we can
immediately verify by integrating out $B_{\mu\nu}^a$, \ie by
substituting the equation of motion of $B_{\mu\nu}^a$,
\begin{equation}
B^a_{\mu\nu} = \frac{g}{2} f^{abc}(A^b_{\mu} A^c_{\nu}) \,.
\end{equation}
Since the BCJ duality is trivially satisfied through four points,
naively one would take the square of this action to obtain a
tree-level action for gravity valid through four points.  However,
since the squaring is with respect to the numerators $n_i$ and not
numerator over propagator, $n_i/s_\alpha$, we need the auxiliary
fields to generate the numerators with the inverse propagators
directly, instead of multiplying and dividing by inverse propagators
afterwards. This implies that the auxiliary fields must become
dynamical (to generate the required propagator) and that their
interactions must contain additional derivatives to produce the
inverse propagator necessary to cancel the propagator.  At four points
this leads to the Lagrangian:
\begin{equation}
 \mathcal{L}_{YM} = \tfrac{1}{2} A^{a\mu}\square A_{\mu}^{a}
  - B^{a\mu\nu\rho}\square B_{\mu\nu\rho}^{a}
  - g f^{abc}(\partial_{\mu} A_{\nu}^{a}
  + \partial^{\rho}B_{\rho\mu\nu}^{a}) A^{b\mu} A^{c\nu}\,,
\end{equation}
where the equation of motion for the auxiliary field $B^a_{\mu\nu\rho}$ becomes
\begin{equation}
 \square B^a_{\mu\nu\rho}= \frac{g}{2} f^{abc} \partial_\mu (A^b_{\nu} A^c_{\rho}) \,.
\end{equation}

We are now ready to construct a gravity action that gives the correct
four-point amplitude. We begin in momentum space, where
 the identification
\begin{equation}
 A_{\mu}(k)\tilde{A}_{\nu}(k)  \rightarrow h_{\mu\nu}(k)
\end{equation}
can be trivially implemented. We first demonstrate how the gravity
action can be derived from the four-point Yang-Mills Lagrangian. We
write the Yang-Mills Lagrangian in momentum space. Since gravity does
not have any color indices, we encode the information of the structure
constants in the anti-symmetrization and cyclicity of the interaction
terms. We drop the coupling constant for now; it can easily be restored
in the final gravity action. We arrive at
\begin{align}
\mathcal{S}_{YM} &\sim \frac{1}{2}\int d^4k_1 d^4k_2 \,\delta^4(k_1+k_2) {k_2^2}
 \Bigl[ A^{\mu}(k_1) A_{\mu}(k_2) -2 B^{\mu\nu\rho}(k_1) B_{\mu\nu\rho}(k_2) \Bigr]\,, \nonumber\\
  &\quad + \int
  d^4k_1d^4k_2d^4k_3\,
  \delta^4(k_1+k_2+k_3) \\
  &\qquad \times P_6 \Bigl\{ \bigl[k_{1\mu}A_{\nu}(k_1) + k_{1}^{\rho}B_{\rho\mu\nu}(k_1)\bigr]A^{\mu}(k_2)A^{\nu}(k_3) \Bigr\} \,,\nonumber
\end{align}
where $P_6$ indicates a sum over all permutations of $\{k_1,k_2,k_3\}$ with the anti-symmetrization signs included. From here, we can read off a gravity action valid through four points:
\begin{align}
  \mathcal{S}_{\rm grav} = \mathcal{S}_{\rm kin} + \mathcal{S}_{\rm int} \,,
\end{align}
with
\begin{align}
  \mathcal{S}_{\rm kin}
 &\sim \frac{1}{4}\int d^4k_1 d^4k_2 \delta^4(k_1+k_2) {k_2^2} \nn\\
      &\quad \times \Bigl[ A^{\mu}(k_1) A_{\mu}(k_2) - 2B^{\mu\nu\rho}(k_1)B_{\mu\nu\rho}(k_2) \Bigr] \nn \\
      &\quad \times \Bigl[ \tilde{A}^{\sigma}(k_1) \tilde{A}_{\sigma}(k_2) - 2\tilde{B}^{\sigma\tau\lambda}(k_1) \tilde{B}_{\sigma\tau\lambda}(k_2) \Bigr]
    \,,\nn \\
  \mathcal{S}_{\rm int} &\sim
    \int d^4k_1d^4k_2d^4k_3
    \delta^4(k_1+k_2+k_3) \nn \\
    &\quad \times P_6 \Bigl\{
	\bigl[k_{1\mu}A_{\nu}(k_1) + k_{1}^{\rho}B_{\rho\mu\nu}(k_1)\bigr]A^{\mu}(k_2)A^{\nu}(k_3)
      \Bigr\} \nn \\
    &\quad \times P_6 \Bigl\{
	\bigl[k_{1\lambda} \tilde{A}_{\sigma}(k_1) + k_{1}^{\tau} \tilde{B}_{\tau\lambda\sigma}(k_1)\bigr]  \tilde{A}^{\lambda}(k_2) \tilde{A}^{\sigma}(k_3)
      \Bigr\} \,.
\end{align}
In extracting the Feynman rules from this action the ``left'' and
``right'' fields each contract independently.  For example, for the
propagators we have,
\begin{align}
&\langle A_\mu (k_1)\tilde A_\rho (k_1)    A_\nu(k_2)
   \tilde A_\sigma(k_2) \rangle={i\eta_{\mu\nu} \eta_{\rho\sigma}
          \over k_1^2 } \delta^4 (k_1 + k_2)\,, \nn \\
& \langle A_\mu (k_1)\tilde B_{\rho\sigma\tau} (k_1) A_\nu(k_2)
    \tilde B_{\eta\kappa\lambda} (k_2) \rangle = -{i\eta_{\mu\nu} \eta_{\rho\eta}
\eta_{\sigma\kappa} \eta_{\tau \lambda}
 \over 2k_1^2 } \delta^4 (k_1 + k_2)  \,.
\end{align}
By construction, this action will give the correct three- and
four-graviton tree-level amplitudes.

We note that one can construct the coordinate-space action by
combining the left-right fields as
\begin{eqnarray}
 A^\mu \tilde{A}^\nu &\rightarrow &h^{\mu\nu}\,, \nn\\
A^\mu \tilde{B}^{\nu\rho\sigma} &\rightarrow& g^{\mu\nu\rho\sigma}\,, \nn\\
B^{\mu\rho\sigma}\tilde{A}^\nu
     &\rightarrow& \tilde{g}^{\mu\rho\sigma\nu}\,, \nn \\
 B^{\mu\rho\sigma}\tilde{B}^{\nu\tau\lambda}
      &\rightarrow&f^{\mu\rho\sigma\nu\tau\lambda}\,,
\end{eqnarray}
where $h^{\mu\nu}$ is the physical field, which includes the graviton, anti-symmetric tensor, and dilaton.
The kinetic terms in $x$ space take the form
\begin{equation}
\mathcal{S}_{\rm kin}=-\frac{1}{2}\int d^Dx
\Bigl[
 h^{\mu\nu}\square h_{\mu\nu}-2g^{\mu\nu\rho\sigma}\square g_{\mu\nu\rho\sigma}-2\tilde{g}^{\mu\nu\rho\sigma}\square \tilde{g}_{\mu\nu\rho\sigma}+4f^{\mu\nu\rho\sigma\tau\lambda}\square f_{\mu\nu\rho\sigma\tau\lambda}\Bigr] \,.
\end{equation}
The interaction terms can similarly be constructed, but we do not
display them here as there are 144 of them.

To move on to five points, we need to introduce a new set of auxiliary
fields to rewrite the non-local terms in a local and cubic form. We
simply give the result:
\begin{align}\label{Lprime5}
  \mathcal{L}'_5 &\rightarrow Y^{a\mu\nu}\square X^{a}_{\mu\nu}
  + D_{(3)}^{a\mu\nu\rho}\square C_{(3)}^{a}\,_{\mu\nu\rho}
  + D_{(4)}^{a\mu\nu\rho\sigma} \square C_{(4)}^{a}\,_{\mu\nu\rho\sigma}\nn \\
    &\quad + g f^{abc} \left( Y^{a\mu\nu} A^{b}_{\mu} A^{c}_{\nu}
	+ \partial_{\mu}D_{(3)}^{a\mu\nu\rho}A^{b}_\nu A^{c}_\rho
	- \tfrac{1}{2} \partial_{\mu} D_{(4)}^{a\mu\nu\rho\sigma} \partial^{\phantom{b}}_{[\nu} A^{b}_{\rho]} A^{c}_{\sigma} \right)\nn \\
    &\quad + g f^{abc} X^{a\mu\nu} \left(
	\tfrac{1}{2} \partial_{\rho} C_{(3)}^{b\rho\sigma}\,_{\mu} \partial^{\phantom{c}}_{[\sigma} A^{c}_{\nu]}
	+ \partial_{\rho} C_{(4)}^{b\rho\sigma}\,_{\nu[\mu} A^{c}_{\sigma]}\right).
\end{align}
Note that these new auxiliary fields do not couple to
$B^{\mu\nu\rho}$.  It is now straight-forward to
transform~(\ref{Lprime5}) to momentum space and, through the squaring
process, obtain a gravity Lagrangian that is valid through five
points.

\subsection{Beyond five points}

As we increase the number of legs we find new violations of
manifest BCJ duality, so we need to add further terms.  We have
constructed a six-point correction to the interactions so that
the Lagrangian generates numerators with manifest BCJ duality.

The
general structure of $\mathcal{L}'_6$ is similar to
that of $\mathcal{L}'_5$;
after relabeling color indices, we can arrange
$\mathcal{L}'_6$ to vanish by two Jacobi identities:
\begin{eqnarray}
  0 &= \left(f^{a_1a_2b}f^{ba_3c} + f^{a_2a_3b}f^{ba_1c} + f^{a_3a_1b}f^{ba_2c}\right) f^{cda_6}f^{da_4a_5} \,, \nn \\
  0 &= f^{a_1a_2b} \left(f^{ba_3c}f^{cda_6} + f^{bdc}f^{ca_6a_3} + f^{ba_6c}f^{ca_3d}\right) f^{da_4a_5}.
\end{eqnarray}
The first of these two color factors is contracted with 59 different
terms having a schematic form\footnote{Momentum conservation can alter these
counts but we give them as an indication of the number of terms involved.}
\begin{equation}
 \frac{1}{\square}(A^{a_1} A^{a_2} A^{a_3})
   \frac{1}{\square}(A^{a_4} A^{a_5}) A^{a_6}\,,
\end{equation}
where the parenthesis indicate which fields the $\frac{1}{\square}$ acts on.
The second color factor contracts with 49 terms of the form
\begin{equation}
  \frac{1}{\square}(A^{a_1} A^{a_2}) A^{a_3}
  \frac{1}{\square}(A^{a_4} A^{a_5}) A^{a_6} \,.
\end{equation}
In each term, there are an additional two partial derivatives in the
numerator acting on the gauge fields. The large number of terms arises
from the many different ways to contract the 8 Lorentz indices. We
have found that the coefficients of these 108 terms depend on 30
distinct free parameters, in addition to the $\beta$ that showed up
at five points (\ref{DFive}).
Thus, there is a 30-parameter family of self-BCJ six-point interactions.

We anticipate that this structure continues to higher orders, with the
addition of new vanishing combinations of terms. We have seen no
indication that the Lagrangian will terminate; for each extra leg that
we add to an amplitude, we will likely need to add more terms to the
action to ensure that the diagrams satisfy the BCJ duality.  A key
outstanding problem is to find a pattern or symmetry that would enable
us to write down the all-order BCJ-corrected action without having to
analyze each $n$-point level at a time.

If the construction of a Lagrangian to all orders succeeds, it would  be a fully off-shell realization of the
BCJ duality at the classical level. It would be interesting to then study non-perturbative phenomena such as instantons using this Lagrangian to see whether BCJ duality
 and the squaring relations can elucidate physics beyond the regime of scattering amplitudes.
Our off-shell construction  suggests that BCJ
duality may also work at loop level.  Of course, one would
need to account for the ghost structure and, more importantly,
demonstrate that the loop amplitudes so constructed do indeed have the
desired duality properties manifest.

\section{A few simple implications}
\label{SimpleImplicationsSection}

\begin{figure}[t]
\centerline{\epsfxsize 3 truein \epsfbox{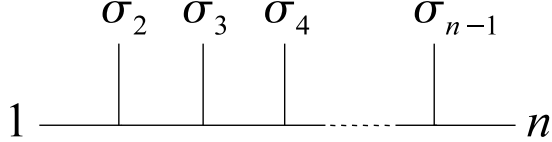}}
\caption[a]{\small A graphical representation of the color basis
  $c_{1, \sigma_2, \ldots, \sigma_{n-1}, n}$
  introduced in ref.~\cite{DDMColor}. Each vertex represents a structure
  constant $\f^{abc}$, while each bond indicates contracted indices
  between the $\f^{abc}$.  This is also precisely the diagram
  associated with the kinematic numerator $n_{1, \sigma_2, \ldots,
    \sigma_{n-1}, n}$.  }
\label{chainFigure}
\end{figure}

In this short section, we point out that the BCJ duality immediately
leads to some novel forms of gauge and gravity amplitudes.  Del Duca,
Dixon and Maltoni~\cite{DDMColor} presented an alternative color
decomposition from the usual one,
\begin{equation}
{\cal A}^\tree_n(1,2,\ldots,n)=
g^{n-2} \sum_{\sigma \in S_{n-2}}
 c_{1, \sigma_2, \ldots, \sigma_{n-1}, n}
 A_n^\tree(1, \sigma_2, \ldots, \sigma_{n-1}, n) \,,
\label{AdjointColor}
\end{equation}
where ${\cal A}^\tree_n$ is the full color-dressed $n$-gluon
amplitude, and the $A_n^\tree$ are the usual color-ordered
partial gauge-theory amplitudes. The sum runs over all permutations
of $n-2$ legs. The color factors are
\begin{equation}
c_{1, \sigma_2, \ldots, \sigma_{n-1}, n} \equiv \f^{a_1 a_{\sigma_2}
  x_1} \f^{x_1 a_{\sigma_3}x_2} \cdots \f^{x_{n-3} a_{\sigma_{n-1}} a_n}\,.
\end{equation}
Diagrammatically, this color factor is associated with
\fig{chainFigure}. This form is derived starting from
\eqn{Anrep} and using color Jacobi identities along with the Kleiss-Kuijf
relations, which are equivalent to the self-anti-symmetry of the
diagrammatic numerator factors.

A simple observation is that when diagram numerators are chosen to
satisfy the BCJ duality, they have precisely the same algebraic
structure as color factors. Thus, we can immediately write a dual
formula decomposing the full amplitude into numerators instead of
color factors:
\begin{equation}
{\cal A}^\tree_n(1,2,\ldots,n)=
g^{n-2} \sum_{\sigma \in S_{n-2}}
n_{1, \sigma_2, \ldots, \sigma_{n-1}, n} \,
 A_n^\scalar
 (1, \sigma_2, \ldots, \sigma_{n-1}, n) \,,
\label{DualAdjoint}
\end{equation}
where $A_n^\scalar$ is a dual partial scalar amplitude with ordered
legs obtained by replacing the gauge-theory numerator factors with
group-theory color factors.  The numerator factors $n_{1, \sigma_2,
  \ldots, \sigma_{n-1}, n}$ are the numerators of the diagrams
displayed in \fig{chainFigure}.  Note that all other numerators
can be expressed as linear combinations of the $n_{1, \sigma_2,
  \ldots, \sigma_{n-1}, n}$ through the duality relations, and the
form~(\ref{DualAdjoint}) for the gauge-theory amplitude makes this
property manifest.  This form is related to an unusual color
decomposition of gauge-theory amplitudes which follows from applying
KLT relations to the low-energy limit of heterotic
strings~\cite{HeteroticKLT}.

Similarly, this immediately gives us a new representation for
graviton amplitudes in terms of gauge-theory amplitudes,
\begin{equation}
{\cal M}^\tree_n(1,2,\ldots,n)=i
\kappa^{n-2} \sum_{\sigma \in S_{n-2}}
n_{1, \sigma_2, \ldots, \sigma_{n-1}, n} \,
 A_n^\tree(1, \sigma_2, \ldots, \sigma_{n-1}, n) \,,
\label{DualGravity}
\end{equation}
where  $A_n^\tree$ is the usual gauge-theory color-ordered amplitude.

\section{Conclusions}

In this paper we investigated consequences of a curious duality
between color and kinematic numerators of gauge-theory
diagrams~\cite{BCJ}.  In particular, using BCFW recursion relations,
we proved that the duality implies that numerators of gravity
amplitudes are just a product of two gauge-theory numerators, as
conjectured in ref.~\cite{BCJ}.  The inductive proof of these
``squaring relations'' makes use of a generalized gauge invariance of
gauge-theory amplitudes, which we use to rearrange the gravity BCFW
recursion relations.  We also explained how the proof extends to other
theories including $\NeqEight$ supergravity as two copies of
$\NeqFour$ super-Yang-Mills theory.  We showed that the squaring
relations even hold in a generalized, asymmetric form, in which only
one set of gauge-theory numerators is required to satisfy the BCJ
duality.  If we assume the duality works at loop level as well, the
unitarity method straightforwardly allows us to conclude that the
squaring relations hold at loop level~\cite{BCJLoops}.

In a complementary approach we described the construction of a
Lagrangian whose Feynman diagrams obey the duality.  Remarkably,
through at least six points, and presumably for any number of points,
the Lagrangian differs from the usual Feynman-gauge Lagrangian by
terms that vanish identically by the color Jacobi identity.  The
extra terms, however, have the effect of shuffling terms between
diagrams to make the duality hold.  These extra higher-point terms
are necessarily non-local, but with the use of auxiliary fields we can
make the Lagrangian local at least through six points.

For the case of $\NeqFour$ super-Yang-Mills theory we took some
initial steps to recast the BCJ duality into a form where both the
duality and supersymmetry are manifest.  It would be interesting to
study this further, and to connect this to the recently
uncovered~\cite{Grassmannian} Grassmannian forms of tree-level
scattering amplitudes in $\NeqFour$ super-Yang-Mills theory.

Another interesting problem would be to see if we can construct a
complete Lagrangian respecting the duality valid to all orders.  The
key open problem for doing so is to find a pattern in the additional
terms that generalizes to higher points.  Although we have constructed
a Lagrangian valid through six points (not presented here), it
contains 108 terms and 30 parameters.  Clearly, we should first
resolve its seeming complexity before attempting to construct
an all-order form.  An intriguing question is whether the form of the
Lagrangian can be fixed by imposing symmetry requirements (prior to
applying color Jacobi identities which make the additional terms
vanish).  If an all-order form of the off-shell gauge-theory
Lagrangian and the corresponding double-copy gravity Lagrangian could
indeed be constructed, then it would be natural to try to find a
mapping between their classical solutions.  Such Lagrangians would
thus lend themselves to addressing non-perturbative implications of
the BCJ duality.

Finally, we note that our partial construction of Lagrangians that
generate diagrams respecting the gauge-theory duality between color
and kinematics and the gravity double-copy property provides new
evidence that these properties may extend to loop level as
well. Indeed, this does appear to be the case, as demonstrated in a
concurrent paper~\cite{BCJLoops}.  We hope that a combined effort of
on-shell methods, Lagrangian approaches, and string theory will shed
further light on the origins, scope, and implications of the BCJ
duality.

\section*{Acknowledgements}
We especially thank J.~J.~M.~Carrasco and H.~Johansson for many
enlightening discussions and for sharing the results of
ref.~\cite{BCJLoops}.  We also thank N.~Arkani-Hamed, F.
Cachazo,  D.~Z.~Freedman and especially H.~Elvang for valuable
discussions.  MK is supported in part by the US National Science
Foundation grant PHY-9802484, and ZB, TD and YH are supported by the
US Department of Energy under contract DE-FG03-91ER40662.


\begin{thebibliography}{99}

\bibitem{WittenTopologicalString}
E.~Witten,
Commun.\ Math.\ Phys.\  {\bf 252}, 189 (2004)
[arXiv:hep-th/0312171].

\bibitem{KLT}
H.~Kawai, D.~C.~Lewellen and S.~H.~H.~Tye,
Nucl.\ Phys.\ B {\bf 269}, 1 (1986).

\bibitem{GravityReview}
Z.~Bern,
Living Rev.\ Rel.\  {\bf 5}, 5 (2002) [gr-qc/0206071].

\bibitem{HeteroticKLT}
Z.~Bern, A.~De Freitas and H.~L.~Wong,
Phys.\ Rev.\ Lett.\  {\bf 84}, 3531 (2000)
[arXiv:hep-th/9912033].

\bibitem{BCJ}
Z.~Bern, J.~J.~M.~Carrasco and H.~Johansson,
Phys.\ Rev.\  D {\bf 78}, 085011 (2008)
[arXiv:0805.3993 [hep-ph]].

\bibitem{Bjerrum1}
N.~E.~J.~Bjerrum-Bohr, P.~H.~Damgaard and P.~Vanhove,
Phys.\ Rev.\ Lett.\  {\bf 103}, 161602 (2009)
[arXiv:0907.1425 [hep-th]];\\
%
S.~Stieberger,
arXiv:0907.2211 [hep-th].

\bibitem{Mafra}
C.~R.~Mafra,
  JHEP {\bf 1001}, 007 (2010)
  [arXiv:0909.5206 [hep-th]].
%

\bibitem{Tye}
  H.~Tye and Y.~Zhang,
  arXiv:1003.1732 [hep-th].

\bibitem{Bjerrum2}
N.~E.~J.~Bjerrum-Bohr, P.~H.~Damgaard, T.~Sondergaard and P.~Vanhove,
  arXiv:1003.2403 [hep-th].

\bibitem{BCFW}
 R.~Britto, F.~Cachazo, B.~Feng and E.~Witten,
  Phys.\ Rev.\ Lett.\  {\bf 94}, 181602 (2005)
  [arXiv:hep-th/0501052].

\bibitem{BCFWGravity}
J.~Bedford, A.~Brandhuber, B.~J.~Spence and G.~Travaglini,
Nucl.\ Phys.\  B {\bf 721}, 98 (2005)
[arXiv:hep-th/0502146];\\
%
F.~Cachazo and P.~Svrcek,
arXiv:hep-th/0502160;\\
%
P.~Benincasa, C.~Boucher-Veronneau and F.~Cachazo,
JHEP {\bf 0711}, 057 (2007)
[arXiv:hep-th/0702032];\\
%
N.~Arkani-Hamed and J.~Kaplan,
JHEP {\bf 0804}, 076 (2008)
[arXiv:0801.2385 [hep-th]];\\
%
A.~Hall,
Phys.\ Rev.\  D {\bf 77}, 124004 (2008)
[arXiv:0803.0215 [hep-th]].

\bibitem{BCJLoops}
Z.~Bern, J.~J.~M.~Carrasco and H.~Johansson, to appear.

\bibitem{UnitarityMethod}
Z.~Bern, L.~J.~Dixon, D.~C.~Dunbar and D.~A.~Kosower,
Nucl.\ Phys.\ B {\bf 425}, 217 (1994)
[arXiv:hep-ph/9403226];
%
Nucl.\ Phys.\ B {\bf 435}, 59 (1995)
[arXiv:hep-ph/9409265];\\
%
 Z.~Bern, L.~J.~Dixon and D.~A.~Kosower,
  JHEP {\bf 0408}, 012 (2004)
  [arXiv:hep-ph/0404293].

\bibitem{BDDPR}
 Z.~Bern, L.~J.~Dixon, D.~C.~Dunbar, M.~Perelstein and J.~S.~Rozowsky,
  Nucl.\ Phys.\  B {\bf 530}, 401 (1998)
  [arXiv:hep-th/9802162].

\bibitem{BernGrant}
Z.~Bern and A.~K.~Grant,
Phys.\ Lett.\  B {\bf 457}, 23 (1999)
[arXiv:hep-th/9904026];
%

\bibitem{Siegel}
W.~Siegel,
  Phys.\ Rev.\  D {\bf 48}, 2826 (1993)
  [arXiv:hep-th/9305073].

\bibitem{DDMColor}
 V.~Del Duca, L.~J.~Dixon and F.~Maltoni,
  Nucl.\ Phys.\  B {\bf 571}, 51 (2000)
  [arXiv:hep-ph/9910563].

\bibitem{Kleiss}
  R.~Kleiss and H.~Kuijf,
  Nucl.\ Phys.\  B {\bf 312}, 616 (1989).

\bibitem{GSW}
M.~B.~Green, J.~H.~Schwarz and E.~Witten,
{\it  Cambridge, Uk: Univ. Pr. ( 1987) 469 P.
 (Cambridge Monographs On Mathematical Physics)}

\bibitem{Risager}
  K.~Risager,
  JHEP {\bf 0512}, 003 (2005)
  [arXiv:hep-th/0508206].

\bibitem{ArkaniHamed2008yf}
  N.~Arkani-Hamed and J.~Kaplan,
  JHEP {\bf 0804}, 076 (2008)
  [arXiv:0801.2385 [hep-th]].

\bibitem{PT}
  S.~J.~Parke and T.~R.~Taylor,
  Phys.\ Rev.\ Lett.\  {\bf 56}, 2459 (1986).

\bibitem{Sondergaard}
  T.~Sondergaard,
  Nucl.\ Phys.\  B {\bf 821}, 417 (2009)
  [arXiv:0903.5453 [hep-th]].

\bibitem{Cheung}
  C.~Cheung,
  arXiv:0808.0504 [hep-th].

\bibitem{Elvang}
  H.~Elvang, D.~Z.~Freedman and M.~Kiermaier,
  JHEP {\bf 0904}, 009 (2009)
  [arXiv:0808.1720 [hep-th]].

\bibitem{Elvang2}
  H.~Elvang, D.~Z.~Freedman and M.~Kiermaier,
  JHEP {\bf 0906}, 068 (2009)
  [arXiv:0811.3624 [hep-th]].

\bibitem{Kiermaier2009yu}
  M.~Kiermaier and S.~G.~Naculich,
  JHEP {\bf 0905}, 072 (2009)
  [arXiv:0903.0377 [hep-th]].

\bibitem{Bianchi2008pu}
  M.~Bianchi, H.~Elvang and D.~Z.~Freedman,
  JHEP {\bf 0809}, 063 (2008)
  [arXiv:0805.0757 [hep-th]].

\bibitem{ArkaniHamed2008gz}
  N.~Arkani-Hamed, F.~Cachazo and J.~Kaplan,
  arXiv:0808.1446 [hep-th].

\bibitem{Brandhuber2008pf}
  A.~Brandhuber, P.~Heslop and G.~Travaglini,
  Phys.\ Rev.\  D {\bf 78}, 125005 (2008)
  [arXiv:0807.4097 [hep-th]].

\bibitem{DdimUnitarity}
Z.~Bern and A.~G.~Morgan,
Nucl.\ Phys.\  B {\bf 467}, 479 (1996)
[arXiv:hep-ph/9511336];\\
Z.~Bern, L.~J.~Dixon, D.~C.~Dunbar and D.~A.~Kosower,
Phys.\ Lett.\  B {\bf 394}, 105 (1997)
[arXiv:hep-th/9611127];\\

\bibitem{Gorsky}
  A.~Gorsky and A.~Rosly,
  JHEP {\bf 0601}, 101 (2006)
  [arXiv:hep-th/0510111];
  \\
 P.~Mansfield,
  JHEP {\bf 0603}, 037 (2006)
  [arXiv:hep-th/0511264];
 \\
  J.~H.~Ettle and T.~R.~Morris,
  JHEP {\bf 0608}, 003 (2006)
  [arXiv:hep-th/0605121];
  \\
 H.~Feng and Y.~t.~Huang,
  JHEP {\bf 0904}, 047 (2009)
  [arXiv:hep-th/0611164].

\bibitem{CSW}
  F.~Cachazo, P.~Svrcek and E.~Witten,
  JHEP {\bf 0409}, 006 (2004)
  [arXiv:hep-th/0403047].

\bibitem{Grassmannian}
N.~Arkani-Hamed, F.~Cachazo, C.~Cheung and J.~Kaplan,
JHEP {\bf 1003}, 020 (2010)
[arXiv:0907.5418 [hep-th]];\\
%
M.~Bullimore, L.~Mason and D.~Skinner,
JHEP {\bf 1003}, 070 (2010)
[0912.0539 [hep-th]].

\end{thebibliography}
\end{document}